\documentclass[pra,aps,twocolumn,showpacs,superscriptaddress,longbibliography,floatfix]{revtex4-1}
\usepackage{graphicx}
\usepackage{amsfonts}
\usepackage{amsmath}
\usepackage{amssymb}
\usepackage{pgf,tikz}
\usepackage{mathrsfs}
\usetikzlibrary{arrows}
\usepackage{mathtools}
\usepackage[colorlinks=true,linkcolor=blue,citecolor=blue,urlcolor=blue]{hyperref}
\usepackage[normalem]{ulem}
\allowdisplaybreaks
\usepackage{soul}
\usepackage{cancel}

\newtheorem{dfn}{Definition}

\newtheorem{thm}{Theorem}

\newtheorem{lem}[thm]{Lemma}

\def\be{\begin{equation}}
\def\ee{\end{equation}}
\usepackage{color}

\definecolor{violeta}{cmyk}{0.07,0.90,0,0.34}

\begin{document}

%%%%%%%%%%%%%%%%%%%%%%%%%%%%%%%%%%%%%%%%%%%%%%%%%%%%%%%%%%%%%%%%%%%

\title{Noncontextual wirings}

%%%%%%%%%%%%%%%%%%%%%%%%%%%%%%%%%%%%%%%%%%%%%%%%%%%%%%%%%%%%%%%%%%%

\author{Barbara Amaral}
%\email{barbaraamaral@gmail.com}
 \affiliation{Departamento de Matem\'atica, Universidade Federal de Ouro Preto,
 Ouro Preto, MG, Brazil}
\affiliation{Departamento de F\'isica e Matem\'atica, CAP - Universidade Federal de S\~ao Jo\~ao del-Rei, 36.420-000, Ouro Branco, MG, Brazil} 
 
\author{Ad\'an Cabello}
%\email{adan@us.es}
 \affiliation{Departamento de F\'{\i}sica
 Aplicada II, Universidad de Sevilla, E-41012 Sevilla, Spain}

\author{Marcelo Terra Cunha}
%\email{tcunha@ime.unicamp.br}
% \affiliation{Departamento de Matem\'atica, Universidade Federal de Minas Gerais,
% Caixa Postal 702, 30123-970, Belo Horizonte, MG, Brazil}
\affiliation{Departamento de Matem\'atica Aplicada, IMECC-Unicamp, 13084-970, Campinas, S\~ao Paulo, Brazil}

\author{Leandro Aolita}
%\email{aolita@if.ufrj.br}
 \affiliation{Instituto de F\'isica, Universidade Federal do Rio de Janeiro, Caixa Postal 68528, Rio de Janeiro, RJ 21941-972, Brazil}
 \affiliation{International Institute of Physics, 
 Federal University of Rio Grande do Norte, 
59070-405 Natal,
	 Brazil}

%%%%%%%%%%%%%%%%%%%%%%%%%%%%%%%%%%%%%%%%%%%%%%%%%%%%%%%%%%%%%%%%%%%

\date{\today}

%First discussion: November, 2015 (Natal)
%This version: October 12, 2016 (Barão Geraldo)

\pacs{03.65.Ta, 03.65.Ud, 02.10.Ox}
%03.65.Ta: Foundations of quantum mechanics; measurement theory
%03.65.Ud: Entanglement and quantum nonlocality
%(e.g. EPR paradox, Bell's inequalities, GHZ states, etc.)
%02.10.Ox: Graph theory

\begin{abstract}
Contextuality is a fundamental feature of quantum theory and is necessary for quantum computation and communication. 
Serious steps have therefore been taken towards a formal framework for contextuality as an operational resource. 
However, the most important component for a resource theory --a concrete, explicit form for the free operations of contextuality-- was still missing.
Here we provide such a component by introducing \emph{noncontextual wirings}: a physically-motivated class of contextuality-free operations with a friendly parametrization. We characterize them completely for the general case of black-box measurement devices with arbitrarily many inputs and outputs. As applications, we show that the
relative entropy of contextuality is a \emph{contextuality monotone} and that maximally contextual boxes that serve as \emph{contextuality bits} exist for a broad class of scenarios. 
Our results complete a unified resource-theoretic framework for contextuality and Bell nonlocality.
\end{abstract}

\maketitle

%%%%%%%%%%%%%%%%%%%%%%%%%%%%%%%%%%%%%%%%%%%%%%%%%%%%%%%%%%%%%%%%%%%
% Intro
%%%%%%%%%%%%%%%%%%%%%%%%%%%%%%%%%%%%%%%%%%%%%%%%%%%%%%%%%%%%%%%%%%%
%
\emph{Introduction.}
Quantum contextuality refers to the impossibility of explaining the statistical predictions of quantum theory in terms of models where the measurement outcomes reveal pre-existent system properties whose values are independent on the context, i.e., on which (or whether) other compatible measurements are jointly performed \cite{Specker60,KS67}. Contextuality can be seen as a generalization of Bell nonlocality \cite{Bell66} to single systems, i.e., without the space-like separation restriction. It thus represents an exotic, intrinsically quantum phenomenon with both fundamental and practical implications. 
Contextuality has received lots of attention over the last decade. On the one hand, it has been experimentally studied in a variety of physical setups \cite{HLBBR06, KZGKGCBR09, ARBC09, LLSLRWZ11, BCAFACTP13}.
On the other one, it has been formally identified as a necessary ingredient for universal quantum computing \cite{Raussendorf13, HWVE14, DGBR14} and a resource for random number certification \cite{UZZWYDDK13}, as well as for several other information-processing tasks in the specific case of space-like separated measurements \cite{Brunner13}.
 
This has motivated considerable interest in \emph{resource theories} of both contextuality \cite{GHHHJKW14, HGJKL15} and Bell nonlocality \cite{GWAN12,V14,GA17}. Resource theories give powerful frameworks for the formal treatment of a physical property as an operational resource, adequate for its characterization, quantification, and manipulation \cite{BG15, CFS16}.
Their central component is a special class of transformations, called the \emph{free operations}, that fulfill the essential requirement of mapping every free (i.e., resourceless) object of the theory into a free object. 
Whereas resource-theoretic approaches for quantum nonlocality are highly developed \cite{Barrett05b, Allcock09, GWAN12, Joshi13, V14, LVN14, GA15, GA17}, the operational framework of contextuality as a resource is still incomplete. In Refs.\ \cite{GHHHJKW14, HGJKL15}, an abstract characterization of the axiomatic structure of a resource theory of contextuality was done. However, a concrete specification of the \emph{free operations of contextuality} was not given. Without an explicit parametrization of a physically-motivated class of free operations, a resource theory significantly loses applicability. 
For instance, in Refs.\ \cite{HGJKL15, GHHHJKW14}, an interesting measure of contextuality, called the \emph{relative entropy of contextuality}, was proposed, but only partial monotonicity under a rather restricted subset of contextuality free operations was shown. Monotonicity (non-increase under the corresponding free operations) is the fundamental requirement for a function to be a valid quantifier of a resource.

Here, we fill this gap by introducing the class of \emph{noncontextual wirings}. These are the natural noncontextuality preserving physical operations at hand in the device-independent scenario of black-box measurement devices, where one does not assume any a-priori knowledge of the state or the observables in question. We derive a friendly analytical expression for generic noncontextual wirings applicable to all \emph{nondisturbing boxes}, so that both quantum and post-quantum boxes are covered. In addition, the framework is versatile in that it allows for transformations between systems with different numbers of inputs and outputs as well as different compatibility constraints. 
Furthermore, for space-like separated measurements, the wirings reduce to 
\emph{local operations assisted by shared randomness}, the canonical free operations of Bell nonlocality \cite{GWAN12,V14,GA17}. 
Hence, the framework constitutes a unified resource theory for both contextuality and Bell nonlocality in their most general forms.
As applications, first, we show that an important quantifier called \emph{relative entropy of contextuality} is \emph{monotonous} under all noncontextual wirings, thus closing a major open problem \cite{HGJKL15, GHHHJKW14}. 
Then, for the broad class of  so-called cycle boxes, we show that contextality bits exists in the strongest possible sense: single boxes from which the entire nondisturbing set can be freely obtained with noncontextual wirings.

%%%%%%%%%%%%%%%%%%%%%%%%%%%%%%%%%%%%%%%%%%%%%%%%%%%%%%%%%%%%%%%%%%%
% Contextuality witnesses and graphs

\emph{Nondisturbing boxes.} We consider a black-box measurement device with 
$b$ buttons (inputs) and $l$ lights (outputs), with $b,l\in\mathbb{N}$, such that 
every time a button is pressed a light is turned on. We assume that the number of lights on is always equal to 
the number of buttons pressed. 
Not all buttons are \emph{compatible}, i.e., can be 
 pressed jointly. Each subset of compatible buttons defines a 
\emph{context}. Let $\mathcal{X}=\left\{ 1, 2, \ldots, b\right\}$
represent the set of buttons. The contexts can be encoded in an
\emph{input compatibility hyper-graph} 
 $\mathcal{I}_\mathcal{X}\coloneqq\big\{\boldsymbol{x}^{(j)}\in\{0,1\}^b\big\}_{j=1,\hdots, |\mathcal{I}_\mathcal{X}|}$, 
 where $|\mathcal{I}_\mathcal{X}|$ is the number of contexts and $j$ labels each context. For each $1\leq j\leq |\mathcal{I}_\mathcal{X}|$, $x^{(j)}_i=0$ stands for ``$i$-th button not pressed for context $j$'' 
and $x^{(j)}_i=1$ for ``$i$-th button pressed for context $j$''. 
For any two strings $\boldsymbol{x}^{(j)},\boldsymbol{x}^{(j')}\in\mathcal{I}_\mathcal{X}$,
we denote by $\boldsymbol{x}^{(j)}\succeq\boldsymbol{x}^{(j')}$ the relationship ``$x^{(j)}_i=0$ implies $x^{(j')}_i=0$, for all $i\in\mathcal{X}$''. 
In other words, $\boldsymbol{x}^{(j)}\succeq\boldsymbol{x}^{(j')}$ means that all the buttons not pressed in $j$ are also not pressed in $j'$ [so that pressing additional buttons from $\boldsymbol{x}^{(j')}$ can lead to
$\boldsymbol{x}^{(j)}$ ].
This defines a partial ordering $\mathcal{I}_\mathcal{X}$, and we say that $j$ is a \emph{maximal context} if, for all $1\leq j'\leq |\mathcal{I}_\mathcal{X}|$, $\boldsymbol{x}^{(j')}\succeq\boldsymbol{x}^{(j)}$ implies $\boldsymbol{x}^{(j')}= \boldsymbol{x}^{(j)}$.

Similarly, not all lights can be jointly turned on. Let $\mathcal{A}=\left\{ 1, 2, \ldots, l\right\}$ be the set of lights. 
For the lights, it is more convenient to work with mutual exclusivity constraints. These can be encoded in an
\emph{output exclusivity hyper-graph} 
 $\mathcal{O}_\mathcal{A}\coloneqq\big\{\boldsymbol{a}^{(i)}\in\{0,1\}^{l}\big\}_{i=1,\hdots, |\mathcal{O}_\mathcal{A}|}$, with $|\mathcal{O}_\mathcal{A}|=b$, where
 $i$ labels each exclusivity hyper-edge (one per button). 
 Each $\boldsymbol{a}^{(i)}\in \mathcal{O}_\mathcal{A}$ encodes the maximal subset of (mutually exclusive) lights associated with button $i\in\mathcal{X}$: $a^{(i)}_k=0$ stands for ``$k$-th light not associated with button $i$'' 
and $a^{(i)}_k=1$ for ``$k$-th light associated with button $i$''.
We denote by $\mathcal{A}_{(i)}\coloneqq\{k\in\mathcal{A}:a^{(i)}_k=1\}$ the subset of lights 
(only one of which can be on per run) associated with button $i\in\mathcal{X}$.
Accordingly, we 
denote by $\mathcal{A}_{(\boldsymbol{x})}\coloneqq\bigcup_{x_i=1,\ i \in \mathcal{X}}\mathcal{A}_{(i)}$ the subset of lights associated with all the buttons
pressed in $\boldsymbol{x}\in\mathcal{I}_\mathcal{X}$. 
In turn, we refer to $\mathcal{X}_{(k)}\coloneqq\{i\in\mathcal{X}:\boldsymbol{a}^{(i)}_k=1\}$ as the subset of buttons
associated with light $k\in\mathcal{A}$. 
We restrict throughout to the case in which only incompatible buttons can have common associated lights. That is, we allow that
$\{i, i'\} \subseteq \mathcal{X}_{(k)}$ only if $x_i\times x_{i'}=0$, for all $\boldsymbol{x}\in\mathcal{I}_\mathcal{X}$. 
Finally, we denote by $\boldsymbol{x}_{(k)}\coloneqq\left(x_i:i \in \mathcal{X}_{(k)}\right)$ the substring of $\boldsymbol{x}$ of buttons
associated with light $k$ and
by $\boldsymbol{a}_{(\boldsymbol{x})}\coloneqq\left(a_k: k\in {\mathcal{A}_{(\boldsymbol{x})}}\right)$ the substring of $\boldsymbol{a}$ 
of lights associated with the buttons pressed in $\boldsymbol{x}$.

For any input hyper-graph $\mathcal{I}_\mathcal{X}$ and output hyper-graph $\mathcal{O}_\mathcal{A}$, 
we consider conditional probability distributions 
\be
\label{eq:def_behavior}
\boldsymbol{P}_{\mathcal{A}|\mathcal{X}}\coloneqq\{p_{\mathcal{A}|\mathcal{X}}(\boldsymbol{a},\boldsymbol{x})\}_{\boldsymbol{a}\in\{0,1\}^{l},\ \boldsymbol{x}\in\mathcal{I}_\mathcal{X}},
\ee
equipped with the property that $p_{\mathcal{A}|\mathcal{X}}$ takes positive values for one, and only one, 
of the lights associated with each pressed button. That is, such that, for each 
$\boldsymbol{x}\in\mathcal{I}_\mathcal{X}$, $p_{\mathcal{A}|\mathcal{X}}(\boldsymbol{a},\boldsymbol{x})\neq0$ 
only if $\|\boldsymbol{a}_{\left(i\right)}\|_\mathrm{h}=1$, with $a_{\left(i\right)}= \left(a_k: k \in \mathcal{A}_{(i)}\right)$
the substring of $\boldsymbol{a}$ of lights associated with button $i$ 
and $\|\boldsymbol{a}_{\left(i\right)}\|_\mathrm{h}$ the Hamming norm (number of ones in) of $\boldsymbol{a}_{\left(i\right)}$, 
for every $i\in\mathcal{X}$ for which $x_i=1$. 
%
%for all strings $\boldsymbol{a}\in\mathcal{O}_\mathcal{A}$ and all strings $\boldsymbol{a}$ with an element $a_j=1$ for some $j \notin \mathcal{A}_{(\boldsymbol{x})}$, 
We refer to any such $\boldsymbol{P}_{\mathcal{A}|\mathcal{X}}$ as a \emph{box behavior} relative to $\mathcal{I}_\mathcal{X}$ and $\mathcal{O}_\mathcal{A}$. 
% Each behavior imposes different button-light consistency constraints between buttons in $\mathcal{X}$ and 
% lights in $\mathcal{A}$, dictating which lights can be turned on upon pressing a given button.
A specially relevant class of behaviors is that of \emph{nondisturbing} ones: $\boldsymbol{P}_{\mathcal{A}|\mathcal{X}}$ is said to be nondisturbing if 
\begin{equation}
\label{eq:def_non_disturbance}
\sum_{a_j:j\notin\mathcal{A}_{(\boldsymbol{x}')}}
p_{\mathcal{A}|\mathcal{X}}(\boldsymbol{a},\boldsymbol{x})=
p_{\mathcal{A}_{(\boldsymbol{x}')}\vert\mathcal{X}'}(\boldsymbol{a}_{(\boldsymbol{x}')},\boldsymbol{x}')\ 
\forall\ \substack{\boldsymbol{x}\in\mathcal{I}_\mathcal{X}:\boldsymbol{x}\succeq\boldsymbol{x}' \\ \boldsymbol{a}_{(\boldsymbol{x}')},\ \boldsymbol{x}'\in\mathcal{I}_\mathcal{X}},
\end{equation} 
where $p_{\mathcal{A}_{(\boldsymbol{x}')}|\mathcal{X}'}$ is a conditional probability
distribution over $\boldsymbol{a}_{(\boldsymbol{x}')}$ given $\boldsymbol{x}'$. 
The nondisturbance condition demands that, whenever two contexts have buttons in common, the marginal distribution over the common buttons is independent of the context. It is thus the analogue of the no-signalling condition in Bell scenarios \cite{Brunner13}. 

With this, we can at last provide a precise formal definition of the general mathematical objects of the resource theory.
Namely, we call every set of input and output hyper-graphs $\mathcal{I}_\mathcal{X}$ and
$\mathcal{O}_\mathcal{A}$, respectively, together with a nondisturbing behavior
$\boldsymbol{P}_{\mathcal{A}|\mathcal{X}}$ relative to them, a \emph{box} 
\be
\label{eq:box_def}
\boldsymbol{B}\coloneqq\{\mathcal{I}_\mathcal{X},\mathcal{O}_\mathcal{A},\boldsymbol{P}_{\mathcal{A}|\mathcal{X}}\}.
\ee
%From this definition, we clearly see now that, for nondisturbing boxes, each output exclusivity hyper-edge must be associated to an input button, as previously anticipated. Otherwise, if there are extra incompatibility constraints among lights consistent with different compatible buttons, the output statistics of one button will in general depend on the context.
We call the set of all such nondisturbing boxes $\mathsf{ND}$. 

In turn, the \emph{free objects} of the theory, i.e. the resourceless ones, are given by 
the class $\mathsf{NC}\subset\mathsf{ND}$ of \emph{noncontextual} (NC) boxes, defined by NC box behaviors.
A behavior $\boldsymbol{P}_{\mathcal{A}|\mathcal{X}}$ is NC if it admits a NC hidden-variable model, i.e., if, for all $\boldsymbol{a}\in\{0,1\}^{|\mathcal{A}|}$ and $\boldsymbol{x}\in\mathcal{I}_\mathcal{X}$,
 \begin{eqnarray}
\label{eq:def_non_context_ala_Bell}
p_{\mathcal{A}|\mathcal{X}}(\boldsymbol{a},\boldsymbol{x})=\sum_{\lambda} p_{\Lambda}(\lambda)\, \prod_{j\in\mathcal{A}}\, D_j (a_j\vert \boldsymbol{x}_{(j)},\lambda),
\end{eqnarray} 
 where $\Lambda$ is the hidden variable, taking the value $\lambda$ with probability $p_{\Lambda}(\lambda)$, and 
 $D_j (a_j\vert \boldsymbol{x}_{(j)},\lambda)\coloneqq \delta\left(a_j,f_j(\boldsymbol{x}_{(j)},\lambda)\right)$, where
 $\delta(a_j,f_j(\boldsymbol{x}_{(j)},\lambda))$, with $\delta$ the Kronecker delta, is the $\lambda$-th NC deterministic response function for
 the $j$-th light given the input substring $\boldsymbol{x}_{(j)}$ of buttons associated with light $j$. 
The function $f_j$ encodes the deterministic assignment of $\boldsymbol{x}_{(j)}$ into $a_j$ for the $\lambda$-th global deterministic strategy. 
In particular, it is such that $f_j(\boldsymbol{0},\lambda)=0$ for all $\lambda$ ($j$-th light is off if no associated button is pressed, i.e., if $\boldsymbol{x}_{(j)}=\boldsymbol{0}$).
Note that, since $\mathcal{X}_{(j)}$ does not depend on the context, $\boldsymbol{x}_{(j)}$ involves always the same buttons. Furthermore, all these buttons are pressed exclusively in different contexts. This explains why $D_j (a_j\vert \boldsymbol{x}_{(j)},\lambda)$ can only generate NC behaviors in Eq.\ \eqref{eq:def_non_context_ala_Bell}. In fact, one can verify that, when the contexts are defined by space-like separated buttons, expression \eqref{eq:def_non_context_ala_Bell} reduces to the usual local hidden-variable models of Bell nonlocality \cite{Brunner13}.
Any box outside $\mathsf{NC}$ is called \emph{contextual}. It is a well known fact that measurements on quantum states
can yield contextual boxes.

%%%%%%%%%%%%%%%%%%%%%%%%%%%%%%%%%%%%%%%%%%%%%%%%%%%%%%%%%%%%%%%%%%%%%

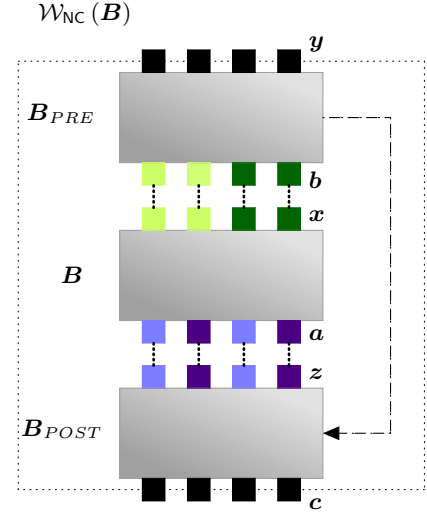
\begin{figure}[t!]
\centering
\definecolor{ubqqys}{rgb}{0.29411764705882354,0.,0.5098039215686274}
\definecolor{xdxdff}{rgb}{0.49019607843137253,0.49019607843137253,1.}
\definecolor{qqwuqq}{rgb}{0.,0.39215686274509803,0.}
\definecolor{ccffww}{rgb}{0.8,1.,0.4}
\definecolor{ttqqqq}{rgb}{0.2,0.,0.}
\definecolor{aqaqaq}{rgb}{0.6274509803921569,0.6274509803921569,0.6274509803921569}
\begin{tikzpicture}[scale=0.15,line cap=round,line join=round,>=triangle 45,x=1.0cm,y=1.0cm]
%\clip(-30.540413354108914,-64.55619875694363) rectangle (134.58282469591467,24.032345843333278);
\fill[fill=aqaqaq, shading = axis,rectangle, left color=aqaqaq, right color=aqaqaq!30!white, shading angle=135] (-4.,8.) -- (-4.,0.) -- (14.,0.) -- (14.,8.) -- cycle;
\fill[color=black,fill=black,fill opacity=1.0] (-2.,10.) -- (-2.,8.) -- (0.,8.) -- (0.,10.) -- cycle;
\fill[fill=black,fill opacity=1.0] (2.,8.) -- (2.,10.) -- (4.,10.) -- (4.,8.) -- cycle;
\fill[fill=black,fill opacity=1.0] (6.,8.) -- (6.,10.) -- (8.,10.) -- (8.,8.) -- cycle;
\fill[fill=black,fill opacity=1.0] (10.,8.) -- (10.,10.) -- (12.,10.) -- (12.,8.) -- cycle;
\fill[color=ccffww,fill=ccffww,fill opacity=1.0] (-2.,0.) -- (-2.,-2.) -- (0.,-2.) -- (0.,0.) -- cycle;
\fill[color=ccffww,fill=ccffww,fill opacity=0.8999999761581421] (2.,0.) -- (2.,-2.) -- (4.,-2.) -- (4.,0.) -- cycle;
\fill[color=qqwuqq,fill=qqwuqq,fill opacity=1.0] (6.,0.) -- (6.,-2.) -- (8.,-2.) -- (8.,0.) -- cycle;
\fill[color=qqwuqq,fill=qqwuqq,fill opacity=1.0] (10.,-2.) -- (10.,0.) -- (12.,0.) -- (12.,-2.) -- cycle;
\fill[fill=aqaqaq,fill opacity=1, shading = axis,rectangle, left color=aqaqaq, right color=aqaqaq!30!white, shading angle=135] (-4.,-6.) -- (14.,-6.) -- (14.,-14.) -- (-4.,-14.) -- cycle;
\fill[color=ccffww,fill=ccffww,fill opacity=1.0] (-2.,-6.) -- (-2.,-4.) -- (0.,-4.) -- (0.,-6.) -- cycle;
\fill[color=ccffww,fill=ccffww,fill opacity=1.0] (2.,-4.) -- (2.,-6.) -- (4.,-6.) -- (4.,-4.) -- cycle;
\fill[color=qqwuqq,fill=qqwuqq,fill opacity=1.0] (6.,-4.) -- (6.,-6.) -- (8.,-6.) -- (8.,-4.) -- cycle;
\fill[color=qqwuqq,fill=qqwuqq,fill opacity=1.0] (10.,-4.) -- (10.,-6.) -- (12.,-6.) -- (12.,-4.) -- cycle;
\fill[color=xdxdff,fill=xdxdff,fill opacity=1.0] (-2.,-14.) -- (-2.,-16.) -- (0.,-16.) -- (0.,-14.) -- cycle;
\fill[color=ubqqys,fill=ubqqys,fill opacity=1.0] (2.,-14.) -- (2.,-16.) -- (4.,-16.) -- (4.,-14.) -- cycle;
\fill[color=xdxdff,fill=xdxdff,fill opacity=1.0] (6.,-14.) -- (6.,-16.) -- (8.,-16.) -- (8.,-14.) -- cycle;
\fill[color=ubqqys,fill=ubqqys,fill opacity=1.0] (10.,-14.) -- (10.,-16.) -- (12.,-16.) -- (12.,-14.) -- cycle;
\fill[fill=aqaqaq,fill opacity=1, shading = rectangle, left color=aqaqaq, right color=aqaqaq!30!white, shading angle=135] (-4.,-20.) -- (-4.,-28.) -- (14.,-28.) -- (14.,-20.) -- cycle;
\fill[color=xdxdff,fill=xdxdff,fill opacity=1.0] (-2.,-20.) -- (-2.,-18.) -- (0.,-18.) -- (0.,-20.) -- cycle;
\fill[color=ubqqys,fill=ubqqys,fill opacity=1.0] (2.,-18.) -- (2.,-20.) -- (4.,-20.) -- (4.,-18.) -- cycle;
\fill[color=xdxdff,fill=xdxdff,fill opacity=1.0] (6.,-18.) -- (6.,-20.) -- (8.,-20.) -- (8.,-18.) -- cycle;
\fill[color=ubqqys,fill=ubqqys,fill opacity=1.0] (10.,-18.) -- (10.,-20.) -- (12.,-20.) -- (12.,-18.) -- cycle;
\fill[fill=black,fill opacity=1.0] (-2.,-28.) -- (-2.,-30.) -- (0.,-30.) -- (0.,-28.) -- cycle;
\fill[fill=black,fill opacity=1.0] (2.,-28.) -- (2.,-30.) -- (4.,-30.) -- (4.,-28.) -- cycle;
\fill[fill=black,fill opacity=1.0] (6.,-28.) -- (6.,-30.) -- (8.,-30.) -- (8.,-28.) -- cycle;
\fill[fill=black,fill opacity=1.0] (10.,-28.) -- (10.,-30.) -- (12.,-30.) -- (12.,-28.) -- cycle;
\draw [color=aqaqaq] (-4.,8.)-- (-4.,0.);
\draw [color=aqaqaq] (-4.,0.)-- (14.,0.);
\draw [color=aqaqaq] (14.,0.)-- (14.,8.);
\draw [color=aqaqaq] (14.,8.)-- (-4.,8.);
\draw [color=black] (-2.,10.)-- (-2.,8.);
\draw [color=black] (-2.,8.)-- (0.,8.);
\draw [color=black] (0.,8.)-- (0.,10.);
\draw [color=black] (0.,10.)-- (-2.,10.);
\draw (2.,8.)-- (2.,10.);
\draw (2.,10.)-- (4.,10.);
\draw (4.,10.)-- (4.,8.);
\draw (4.,8.)-- (2.,8.);
\draw (6.,8.)-- (6.,10.);
\draw (6.,10.)-- (8.,10.);
\draw (8.,10.)-- (8.,8.);
\draw (8.,8.)-- (6.,8.);
\draw (10.,8.)-- (10.,10.);
\draw (10.,10.)-- (12.,10.);
\draw (12.,10.)-- (12.,8.);
\draw (12.,8.)-- (10.,8.);
\draw [color=ccffww] (-2.,0.)-- (-2.,-2.);
\draw [color=ccffww] (-2.,-2.)-- (0.,-2.);
\draw [color=ccffww] (0.,-2.)-- (0.,0.);
\draw [color=ccffww] (0.,0.)-- (-2.,0.);
\draw [color=ccffww] (2.,0.)-- (2.,-2.);
\draw [color=ccffww] (2.,-2.)-- (4.,-2.);
\draw [color=ccffww] (4.,-2.)-- (4.,0.);
\draw [color=ccffww] (4.,0.)-- (2.,0.);
\draw [color=qqwuqq] (6.,0.)-- (6.,-2.);
\draw [color=qqwuqq] (6.,-2.)-- (8.,-2.);
\draw [color=qqwuqq] (8.,-2.)-- (8.,0.);
\draw [color=qqwuqq] (8.,0.)-- (6.,0.);
\draw [color=qqwuqq] (10.,-2.)-- (10.,0.);
\draw [color=qqwuqq] (10.,0.)-- (12.,0.);
\draw [color=qqwuqq] (12.,0.)-- (12.,-2.);
\draw [color=qqwuqq] (12.,-2.)-- (10.,-2.);
\draw [color=aqaqaq] (-4.,-6.)-- (14.,-6.);
\draw [color=aqaqaq] (14.,-6.)-- (14.,-14.);
\draw [color=aqaqaq] (14.,-14.)-- (-4.,-14.);
\draw [color=aqaqaq] (-4.,-14.)-- (-4.,-6.);
\draw [color=ccffww] (-2.,-6.)-- (-2.,-4.);
\draw [color=ccffww] (-2.,-4.)-- (0.,-4.);
\draw [color=ccffww] (0.,-4.)-- (0.,-6.);
\draw [color=ccffww] (0.,-6.)-- (-2.,-6.);
\draw [color=ccffww] (2.,-4.)-- (2.,-6.);
\draw [color=ccffww] (2.,-6.)-- (4.,-6.);
\draw [color=ccffww] (4.,-6.)-- (4.,-4.);
\draw [color=ccffww] (4.,-4.)-- (2.,-4.);
\draw [color=qqwuqq] (6.,-4.)-- (6.,-6.);
\draw [color=qqwuqq] (6.,-6.)-- (8.,-6.);
\draw [color=qqwuqq] (8.,-6.)-- (8.,-4.);
\draw [color=qqwuqq] (8.,-4.)-- (6.,-4.);
\draw [color=qqwuqq] (10.,-4.)-- (10.,-6.);
\draw [color=qqwuqq] (10.,-6.)-- (12.,-6.);
\draw [color=qqwuqq] (12.,-6.)-- (12.,-4.);
\draw [color=qqwuqq] (12.,-4.)-- (10.,-4.);
\draw [color=xdxdff] (-2.,-14.)-- (-2.,-16.);
\draw [color=xdxdff] (-2.,-16.)-- (0.,-16.);
\draw [color=xdxdff] (0.,-16.)-- (0.,-14.);
\draw [color=xdxdff] (0.,-14.)-- (-2.,-14.);
\draw [color=ubqqys] (2.,-14.)-- (2.,-16.);
\draw [color=ubqqys] (2.,-16.)-- (4.,-16.);
\draw [color=ubqqys] (4.,-16.)-- (4.,-14.);
\draw [color=ubqqys] (4.,-14.)-- (2.,-14.);
\draw [color=xdxdff] (6.,-14.)-- (6.,-16.);
\draw [color=xdxdff] (6.,-16.)-- (8.,-16.);
\draw [color=xdxdff] (8.,-16.)-- (8.,-14.);
\draw [color=xdxdff] (8.,-14.)-- (6.,-14.);
\draw [color=ubqqys] (10.,-14.)-- (10.,-16.);
\draw [color=ubqqys] (10.,-16.)-- (12.,-16.);
\draw [color=ubqqys] (12.,-16.)-- (12.,-14.);
\draw [color=ubqqys] (12.,-14.)-- (10.,-14.);
\draw [color=aqaqaq] (-4.,-20.)-- (-4.,-28.);
\draw [color=aqaqaq] (-4.,-28.)-- (14.,-28.);
\draw [color=aqaqaq] (14.,-28.)-- (14.,-20.);
\draw [color=aqaqaq] (14.,-20.)-- (-4.,-20.);
\draw [color=xdxdff] (-2.,-20.)-- (-2.,-18.);
\draw [color=xdxdff] (-2.,-18.)-- (0.,-18.);
\draw [color=xdxdff] (0.,-18.)-- (0.,-20.);
\draw [color=xdxdff] (0.,-20.)-- (-2.,-20.);
\draw [color=ubqqys] (2.,-18.)-- (2.,-20.);
\draw [color=ubqqys] (2.,-20.)-- (4.,-20.);
\draw [color=ubqqys] (4.,-20.)-- (4.,-18.);
\draw [color=ubqqys] (4.,-18.)-- (2.,-18.);
\draw [color=xdxdff] (6.,-18.)-- (6.,-20.);
\draw [color=xdxdff] (6.,-20.)-- (8.,-20.);
\draw [color=xdxdff] (8.,-20.)-- (8.,-18.);
\draw [color=xdxdff] (8.,-18.)-- (6.,-18.);
\draw [color=ubqqys] (10.,-18.)-- (10.,-20.);
\draw [color=ubqqys] (10.,-20.)-- (12.,-20.);
\draw [color=ubqqys] (12.,-20.)-- (12.,-18.);
\draw [color=ubqqys] (12.,-18.)-- (10.,-18.);
\draw (-2.,-28.)-- (-2.,-30.);
\draw (-2.,-30.)-- (0.,-30.);
\draw (0.,-30.)-- (0.,-28.);
\draw (0.,-28.)-- (-2.,-28.);
\draw (2.,-28.)-- (2.,-30.);
\draw (2.,-30.)-- (4.,-30.);
\draw (4.,-30.)-- (4.,-28.);
\draw (4.,-28.)-- (2.,-28.);
\draw (6.,-28.)-- (6.,-30.);
\draw (6.,-30.)-- (8.,-30.);
\draw (8.,-30.)-- (8.,-28.);
\draw (8.,-28.)-- (6.,-28.);
\draw (10.,-28.)-- (10.,-30.);
\draw (10.,-30.)-- (12.,-30.);
\draw (12.,-30.)-- (12.,-28.);
\draw (12.,-28.)-- (10.,-28.);
\draw [line width=1.pt,dash pattern=on 0.5pt off 1.5pt] (-1.,-4.)-- (-1.,-2.);
\draw [line width=1.pt,dash pattern=on 0.5pt off 1.5pt] (3.,-4.)-- (3.,-2.);
\draw [line width=1.pt,dash pattern=on 0.5pt off 1.5pt] (7.,-4.)-- (7.,-2.);
\draw [line width=1.pt,dash pattern=on 0.5pt off 1.5pt] (11.,-4.)-- (11.,-2.);
\draw [line width=1.pt,dash pattern=on 0.5pt off 1.5pt] (-1.,-18.)-- (-1.,-16.);
\draw [line width=1.pt,dash pattern=on 0.5pt off 1.5pt] (3.,-18.)-- (3.,-16.);
\draw [line width=1.pt,dash pattern=on 0.5pt off 1.5pt] (7.,-18.)-- (7.,-16.);
\draw [line width=1.pt,dash pattern=on 0.5pt off 1.5pt] (11.,-18.)-- (11.,-16.);
\draw (-13,6) node[anchor=north west] {$\boldsymbol{B}_{PRE}$};
\draw (-10,-8.5) node[anchor=north west] {$\boldsymbol{B}$};
\draw (-13.5,-22) node[anchor=north west] {$\boldsymbol{B}_{POST}$};
\draw (12,12) node[anchor=north west] {$\boldsymbol{y}$};
\draw (12,0) node[anchor=north west] {$\boldsymbol{b}$};
\draw (12,-3.5) node[anchor=north west] {$\boldsymbol{x}$};
\draw (12,-14) node[anchor=north west] {$\boldsymbol{a}$};
\draw (12,-29) node[anchor=north west] {$\boldsymbol{c}$};
\draw (12,-17.5) node[anchor=north west] {$\boldsymbol{z}$};
\draw [dotted] (-13.,9.)-- (23.,9.);
\draw [dotted] (23.,9.)-- (23.,-29.);
\draw [dotted] (23.,-29.)-- (-13.,-29.);
\draw [dotted] (-13.,-29.)-- (-13.,9.);
\draw [->,dash pattern=on 0.5 pt off 2pt on 4pt off 0.5pt] (20.,-24.) -- (14.,-24.);
\draw [dash pattern=on 0.5 pt off 2pt on 4pt off 0.5pt] (14.,4.)-- (20.,4.);
\draw [dash pattern=on 0.5 pt off 2pt on 4pt off 0.5pt] (20.,4.)-- (20.,-24.);
\draw (-12,15) node[anchor=north west] {$\mathcal{W}_{\mathsf{NC}}\left(\boldsymbol{B}\right)$};
\end{tikzpicture}
 \caption{A noncontextual wiring $\mathcal{W}_\mathsf{NC}$ with respect to pre- and post-processing boxes $\boldsymbol{B}_\mathrm{PRE}$ and $\boldsymbol{B}_\mathrm{POST}$, respectively,  maping an initial box $\boldsymbol{B}$ into a final box $\mathcal{W}_\mathsf{NC}(\boldsymbol{B})$. 
 The buttons and lights of $\mathcal{W}_\mathsf{NC}(\boldsymbol{B})$ are given by the buttons of $\boldsymbol{B}_\mathrm{PRE}$ and the lights of $\boldsymbol{B}_\mathrm{POST}$, respectively.
Only the lights (buttons) of $\boldsymbol{B}$ of the same colour can be on (pressed) at the same time. 
The behavior of $\boldsymbol{B}_\mathrm{POST}$ is causally influenced by $\boldsymbol{B}_\mathrm{PRE}$, but in a restricted way such that the statistics of each output light of $\boldsymbol{B}_\mathrm{POST}$ depends only on the buttons and lights of $\boldsymbol{B}_\mathrm{PRE}$ that are associated with it (see text). As a result, if $\boldsymbol{B}$ is noncontextual so is $\mathcal{W}_\mathsf{NC}(\boldsymbol{B})$.
 \label{fig:wirings}
}
\end{figure}

\emph{Contextuality-free operations.} We consider compositions of the initial box $\boldsymbol{B}$ with a 
pre-processing box 
\be
\label{eq:pre_box_def}
\boldsymbol{B}_\mathrm{PRE}\coloneqq\{\mathcal{I}_{\mathcal{Y}},\mathcal{O}_\mathcal{B}, %\mathcal{Y}_{\mathcal{B}},
\boldsymbol{P}_{\mathcal{B}|\mathcal{Y}}\}\in\mathsf{NC}
\ee
 and a $\left(\boldsymbol{b}, \boldsymbol{y} \right)$-dependent post-processing box
\be
\label{eq:pos_box_def}
\boldsymbol{B}_\mathrm{POST}\left(\boldsymbol{b}, \boldsymbol{y} \right) \coloneqq\{\mathcal{I}_{\mathcal{Z}},\mathcal{O}_\mathcal{C},%\mathcal{Z}_{\mathcal{C}},
\boldsymbol{P}_{\mathcal{C}|\mathcal{Z}, \boldsymbol{y}, \boldsymbol{b}}\}\in\mathsf{NC},
\ee
for all $\boldsymbol{b}\in\{0,1\}^{|\mathcal{B}|}$ and $\boldsymbol{y}\in\mathcal{I}_\mathcal{Y}$, as shown in Fig.\ \ref{fig:wirings}. $\mathcal{Y}$ and $\mathcal{B}$ are, respectively, the sets of buttons and lights of $\boldsymbol{B}_\mathrm{PRE}$, and $\mathcal{Z}$ and $\mathcal{C}$ those of $\boldsymbol{B}_\mathrm{POST}\left(\boldsymbol{b}, \boldsymbol{y} \right)$.
For the composition to be possible, we demand that the set 
of allowed outputs [inputs] of $\boldsymbol{B}_\mathrm{PRE}$ [$\boldsymbol{B}_\mathrm{POST}\left(\boldsymbol{b}, \boldsymbol{y} \right)$] is a subset of the set of allowed inputs [outputs] of $\boldsymbol{B}$, i.e., $\overline{\mathcal{O}}_{\mathcal{B}}\subseteq\mathcal{I}_{\mathcal{X}}$ and $\overline{\mathcal{O}}_{\mathcal{A}}\subseteq\mathcal{I}_{\mathcal{Z}}$.
Here, we have introduced the \emph{complementary hyper-graph} $\overline{\mathcal{O}}_{\mathcal{A}}$ to $\mathcal{O}_{\mathcal{A}}$, given by all the output strings with at most one light on per output exclusivity hyper-edge of $\mathcal{O}_{\mathcal{A}}$:
 $\overline{\mathcal{O}}_\mathcal{A}\coloneqq\{\boldsymbol{a}\in\{0,1\}^{|\mathcal{A}|}\colon \forall\ \boldsymbol{a}'
\in\mathcal{O}_\mathcal{A},\ \sum_{k\in\mathcal{A}: a'_k=1} a_k\leq1\}$; and similarly for $\overline{\mathcal{O}}_\mathcal{B}$.
 Thus, $\overline{\mathcal{O}}_\mathcal{A}$ ($\overline{\mathcal{O}}_\mathcal{B}$) gives the combinations of lights on that do not violate any of the exclusivity constraints in $\mathcal{O}_\mathcal{A}$ ($\mathcal{O}_\mathcal{B}$).
 
Moreover, we allow $\boldsymbol{P}_{\mathcal{C}|\mathcal{Z}, \boldsymbol{y}, \boldsymbol{b}}$ to have only a restricted dependence on $\left(\boldsymbol{b}, \boldsymbol{y} \right)$, in such a way that each output light of the post-processing box is causally influenced only by the inputs and outputs of the pre-processing box that are associated with it. That is, we demand that, for all $\boldsymbol{b}\in\{0,1\}^{|\mathcal{B}|}$, $\boldsymbol{c}\in\{0,1\}^{|\mathcal{C}|}$, $\boldsymbol{y}\in\mathcal{I}_\mathcal{Y}$, and $\boldsymbol{z}\in\mathcal{I}_\mathcal{Z}$,
\begin{equation}
\label{eq:def_post_box}
p_{\mathcal{C}|\mathcal{Z}, \boldsymbol{b}, \boldsymbol{y} }(\boldsymbol{c},\boldsymbol{z})=
\sum_{\phi} p_{\Phi}(\phi)\, \prod_{j\in\mathcal{C}}\, D_j (c_j\vert \boldsymbol{z}_{(j)}, \boldsymbol{b}_{[j]}, \boldsymbol{y}_{[j]},\phi),
\end{equation} 
with $ D_j (c_j\vert \boldsymbol{z}_{(j)}, \boldsymbol{b}_{[j]}, \boldsymbol{y}_{[j]},\phi)$ defined analogously to $D_j (a_j\vert \boldsymbol{x}_{(j)},\lambda)$ in Eq.\ \eqref{eq:def_non_context_ala_Bell}.
As before, $\boldsymbol{z}_{(j)}$ is the substring of $\boldsymbol{z}$ associated with light $j\in\mathcal{C}$, corresponding to the subset $\mathcal{Z}_{(j)}\in\mathcal{Z}$ (of all incompatible buttons).
In turn, $\boldsymbol{b}_{[j]}$ and $\boldsymbol{y}_{[j]}$, corresponding $\mathcal{X}_{[j]}\in\mathcal{X}$ and $\mathcal{Y}_{[j]}\in\mathcal{Y}$, respectively, are the substrings of $\boldsymbol{b}$ and $\boldsymbol{y}$ associated with the buttons in $\mathcal{Z}_{(j)}$. More precisely, with $\mathcal{X}_{[j]}\coloneqq \{ \mathcal{X}_{(i)}: i \in \mathcal{Z}_{(j)}\}$ and $\mathcal{Y}_{[j]}\coloneqq \{ \mathcal{Y}_{(i)}: i \in \mathcal{X}_{[j]}\}$, we explicitly write $\boldsymbol{b}_{[j]}\coloneqq\{b_l :l \in \mathcal{X}_{[j]}\}$ and $\boldsymbol{y}_{[j]}\coloneqq\{y_l :l \in \mathcal{Y}_{[j]}\}$. In App.\ \ref{sec:Append0}, we show that, for all $j\in\mathcal{C}$, both $\mathcal{X}_{[j]}\in\mathcal{X}$ and $\mathcal{Y}_{[j]}\in\mathcal{Y}$ are composed of mutually incompatible buttons according to $\mathcal{I}_\mathcal{X}$ and $\mathcal{I}_\mathcal{Y}$, respectively. This is crucial for the composition not to create contextuality.

With this, we are now in a good position to introduce the free operations of contextuality. 
\begin{dfn}[Noncontextual wirings]
\label{def_post_box}
We define the \emph{noncontextual wiring} with respect to pre- and post-processing boxes of Eqs.\ \eqref{eq:pre_box_def} and \eqref{eq:pos_box_def}, respectively, as the linear map $\mathcal{W}_\mathsf{NC}$ that takes any 
initial box $\boldsymbol{B}\in\mathsf{ND}$, given by Eq.\ \eqref{eq:box_def}, into a final box $\boldsymbol{B}_\mathrm{f}\coloneqq\mathcal{W}_\mathsf{NC}(\boldsymbol{B})$ with $b_\mathrm{f}\coloneqq|\mathcal{Y}|$ buttons and $l_\mathrm{f}\coloneqq|\mathcal{C}|$ lights, with
\be
\label{eq:final_box_def}
\mathcal{W}_\mathsf{NC}(\boldsymbol{B})\coloneqq\{\mathcal{I}_{\mathcal{Y}},\mathcal{O}_{\mathcal{C}}, %\mathcal{Y}_{\mathcal{C}},
\boldsymbol{P}_{\mathcal{C}|\mathcal{Y}}\},
\ee
where %$\mathcal{Y}_{\mathcal{C}}\coloneqq\left\{\mathcal{Y}_{[j]}\right\}_{j \in \mathcal{C}}$ is the final the button-light consistency set and 
$\boldsymbol{P}_{\mathcal{C}|\mathcal{Y}}$ is the final behavior, given by
\begin{align}
 \label{eqwiring} 
 \nonumber
 &p_{\mathcal{C}|\mathcal{Y}}(\boldsymbol{c},\boldsymbol{y}) =\\
 &\sum_{\substack{\boldsymbol{\alpha}\in\{0,1\}^{|\mathcal{A}|} \\ \boldsymbol{\beta}\in\{0,1\}^{|\mathcal{B}|}}} p_{\mathcal{C}|\mathcal{Z}, \boldsymbol{\beta}, \boldsymbol{y} }(\boldsymbol{c},\boldsymbol{\alpha}) \, p_{\mathcal{A}|\mathcal{X}}(\boldsymbol{\alpha},\boldsymbol{\beta})\, p_{\mathcal{B}|\mathcal{Y}}(\boldsymbol{\beta},\boldsymbol{y}),
\end{align}
for all $\boldsymbol{c}\in\{0,1\}^{|\mathcal{C}|}$ and $\boldsymbol{y}\in\mathcal{I}_\mathcal{Y}$. We denote the class of all such wirings by $\mathsf{NCW}$.
\end{dfn}
%

%%%%%%%%%%%%%%%%%%%%%%%%%%%%%%%%%%%%%%%%%%%%%%%%%%%%%%%%%%%%%%%%%%%%%

%%%%%%%%%%%%%%%%%%%%%%%%%%%%%%%%%%%%%%%%%%%%%%%%%%%%%%%%%%%%%%%%%%%%%

Self-consistency of the theory requires that the class $\mathsf{NCW}$ satisfies the following property, proven in App.\ \ref{sec:Append1}.
\begin{lem}[Nondisturbance preservation] 
\label{Lem:ND}
The class of boxes $\mathsf{ND}$ is closed under all wirings in $\mathsf{NCW}$.
\end{lem}
 
More importantly, to give valid free operations, $\mathsf{NCW}$ must fulfill the following requirement, proven in App.\ \ref{sec:app_2}.
\begin{thm}[Noncontextuality preservation]
\label{teoncpreservation}
The class of boxes $\mathsf{NC}$ is closed under all wirings in $\mathsf{NCW}$.
 \end{thm}
Intuitively, this is connected to the fact that the composition of any three independent boxes, each one given by a probabilistic mixture of NC deterministic assignments, yields also another such a mixture (with three independent hidden variables). $\mathsf{NCW}$ is more general because the pre- and post-processing boxes are not independent. However, the restricted dependence in Eq.\ \eqref{eq:def_post_box} enables noncontextuality preservation (see App.\ \ref{sec:app_2}).

%%%%%%%%%%%%%%%%%%%%%%%%%%%%%%%%%%%%%%%%%%%%%%%%%%%%%%%%%%%%%%%%%%%%%

\emph{Contextuality monotones.} 
In Ref.\ \cite{GHHHJKW14}, a measure of contextuality called the relative entropy of contextuality $R_\mathrm{C}$ was introduced. For an arbitrary box 
 $\boldsymbol{B}\in\mathsf{ND}$, 
\begin{equation}
\label{eqcontextmonotone}
R_\mathrm{C} (\boldsymbol{B}) \coloneqq \min_{\boldsymbol{B}^{*} \in \mathsf{NC}} S\left(\boldsymbol{B}\| \boldsymbol{B}^{*}\right).
\end{equation}
$S\left(\boldsymbol{B}\| \boldsymbol{B}^{*}\right)$ is the relative entropy of $\boldsymbol{B}$ with respect to $\boldsymbol{B}^{*}$ (defined precisely in App.\ \ref{sec:monotonicity}),
which measures the distinguishability of $\boldsymbol{B}$ from $\boldsymbol{B}^{*}$ in a broad class of scenarios \cite{GA17}. Hence, $R_\mathrm{C} (\boldsymbol{B})$ quantifies the distinguishability of $\boldsymbol{B}$ from its closest (with respect to $S$) noncontextual box $\boldsymbol{B}^{*}$, providing a direct generalisation to contextuality of the statistical strength of Bell nonlocality proofs \cite{vDGG05}. 

The essential requirement for a function to be a valid measure of a resource is that it is monotonous (i.e., non-increasing) under the corresponding free operations. In Ref.\ \cite{GHHHJKW14}, the authors show, for all quantum boxes, monotonicity of $R_\mathrm{C}$ under probabilistic mixtures of independent channels on each quantum observable (each context). This corresponds to a highly restricted subset of $\mathsf{NCW}$ \cite{comment}. Here, we show monotonicity of $R_\mathrm{C}$ under the whole class $\mathsf{NCW}$ and for all boxes $\boldsymbol{B}\in\mathsf{ND}$.

%%%%%%%%%%%%%%%%%%%%%%%%%%%%%%%%%%%%%%%%%%%%%%%%%%%%%%%%%%%%%%%%%%%%%

\begin{lem}[Monotonicity of $R_\mathrm{C}$]
\label{Lem:monotonicity}
Let $\boldsymbol{B}\in\mathsf{ND}$. Then, $R_\mathrm{C}\left(\mathcal{W}_\mathsf{NC}(\boldsymbol{B})\right) \leq R_\mathrm{C}\left(\boldsymbol{B}\right)$ for all $\mathcal{W}_\mathsf{NC}\in\mathsf{NCW}$.
%and 
 %$X_{u}\left(\boldsymbol{B}\right) \leq X_{u}\left(\boldsymbol{B}_0\right).$
 \end{lem}
 The proof relies explicitly on the parametrization of $\mathsf{NCW}$ provided by Eq.\ \eqref{eqwiring}. It can be found in App.\ \ref{sec:monotonicity}. 

%%%%%%%%%%%%%%%%%%%%%%%%%%%%%%%%%%%%%%%%%%%%%%%%%%%%%%%%%%%%%%%%%%%%%

 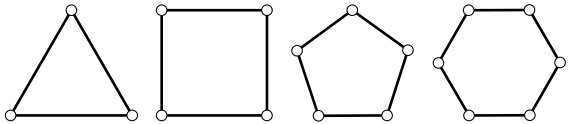
\begin{figure}[t!]
 \definecolor{ffffff}{rgb}{1.,1.,1.}
\begin{tikzpicture}[scale=0.04,line cap=round,line join=round,>=triangle 45,x=1.0cm,y=1.0cm]
%\clip(-23.04291050665253,-69.70740967688049) rectangle (166.21278486725507,31.828187665311233);
%\fill(0.,15.) -- (-20.205022216871633,-20.) -- (20.208378024019535,-19.99806252383982) -- cycle;
%\fill(30.,15.) -- (30.,-20.) -- (65.,-20.) -- (65.,15.) -- cycle;
%\fill(93.36506058891567,15.) -- (75.,1.5191024189244082) -- (82.145979665289,-20.112934578299623) -- (104.92749857026885,-20.001371107403696) -- (111.86127190360571,1.6996159067369216) -- cycle;
%\fill(132.0443869420628,15.) -- (122.,-2.542198094237202) -- (132.16979571679698,-20.011991398622882) -- (152.38397837565677,-19.939586608771364) -- (162.42836531771957,-2.3973885145341702) -- (152.2585696009226,15.072404789851511) -- cycle;
\draw [line width=1pt] (0.,15.)-- (-20.205022216871633,-20.);
\draw [line width=1pt] (-20.205022216871633,-20.)-- (20.208378024019535,-19.99806252383982);
\draw [line width=1pt] (0.,15.)-- (20.208378024019535,-19.99806252383982);
\draw [line width=1pt] (30.,15.)-- (30.,-20.);
\draw [line width=1pt] (65.,-20.)-- (65.,15.);
\draw [line width=1pt] (65.,15.)-- (30.,15.);
\draw [line width=1pt]  (93.36506058891567,15.)-- (75.,1.5191024189244082);
\draw [line width=1pt] (75.,1.5191024189244082)-- (82.145979665289,-20.112934578299623);
\draw [line width=1pt] (82.145979665289,-20.112934578299623)-- (104.92749857026885,-20.001371107403696);
\draw [line width=1pt] (104.92749857026885,-20.001371107403696)-- (111.86127190360571,1.6996159067369216);
\draw [line width=1pt] (111.86127190360571,1.6996159067369216)-- (93.36506058891567,15.);
\draw [line width=1pt] (132.0443869420628,15.)-- (122.,-2.542198094237202);
\draw [line width=1pt] (122.,-2.542198094237202)-- (132.16979571679698,-20.011991398622882);
\draw [line width=1pt] (152.38397837565677,-19.939586608771364)-- (162.42836531771957,-2.3973885145341702);
\draw [line width=1pt] (162.42836531771957,-2.3973885145341702)-- (152.2585696009226,15.072404789851511);
\draw [line width=1pt] (152.2585696009226,15.072404789851511)-- (132.0443869420628,15.);
\draw [line width=1pt]  (30.,-20.)-- (65.,-20.);
\draw [line width=1pt] (132.16979571679698,-20.011991398622882)-- (152.38397837565677,-19.939586608771364);
\draw [fill=ffffff] (0.,15.) circle (50pt);
\draw [fill=ffffff] (-20.205022216871633,-20.) circle (50pt);
\draw [fill=ffffff] (20.208378024019535,-19.998062523839824) circle (50pt);
\draw [fill=ffffff] (30.,15.) circle (50pt);
\draw [fill=ffffff] (30.,-20.) circle (50pt);
\draw [fill=ffffff] (65.,-20.) circle (50pt);
\draw [fill=ffffff] (65.,15.) circle (50pt);
\draw [fill=ffffff] (75.,1.5191024189244082) circle (50pt);
\draw [fill=ffffff] (93.36506058891567,15.) circle (50pt);
\draw [fill=ffffff] (111.86127190360571,1.6996159067369216) circle (50pt);
\draw [fill=ffffff] (82.14597966528899,-20.11293457829962) circle (50pt);
\draw [fill=ffffff] (104.92749857026887,-20.00137110740369) circle (50pt);
\draw [fill=ffffff] (132.0443869420628,15.) circle (50pt);
\draw [fill=ffffff] (122.,-2.542198094237202) circle (50pt);
\draw [fill=ffffff] (132.16979571679698,-20.011991398622882) circle (50pt);
\draw [fill=ffffff] (152.38397837565677,-19.939586608771364) circle (50pt);
\draw [fill=ffffff] (162.42836531771957,-2.3973885145341702) circle (50pt);
\draw [fill=ffffff] (152.2585696009226,15.072404789851511) circle (50pt);
4
\end{tikzpicture}
 \caption{$b$-cycle graphs $C_b$ for $b=3$, 4, 5, and 6 buttons. A $b$-cycle box is such that the union of all hyper-edges 
 in $\mathcal{I}_{\mathcal{X}}$ equals $C_b$ and each input button has its own pair of associated output lights. 
 For  even $b$, the class is also intimately connected to the well-known chained inequalities of Bell nonlocality \cite{BC90}. 
 It includes the Clauser-Horne-Shimony-Holt (CHSH) scenario \cite{CHSH69}, where the inputs define the square $C_4$, and the
 Klyachko-Can-Binicio\v{g}lu-Shumovsky one \cite{KCBS08}, where the inputs form the pentagon $C_5$. 
 For any $b\geq3$, there exist contextuality bits, i.e., maximally contextual $b$-cycle boxes from which all other $b$-cycle boxes can be obtained for free (see text). 
 \label{figncycle}}
 \end{figure}

%%%%%%%%%%%%%%%%%%%%%%%%%%%%%%%%%%%%%%%%%%%%%%%%%%%%%%%%%%%%%%%%%%%%%

\emph{Contextuality bits.}
The operational framework developed allows us to study contextuality interconversions. A natural question is whether there exists a box from which all boxes, for fixed input and output hyper-graphs, can be obtained \emph{for free} (i.e., through noncontextual wirings). This is intimately connected to quantification: such a superior box can be taken as \emph{unit of contextuality}, or \emph{contextuality bit}, yielding a natural and unambiguous (measure-independent) definition of maximally contextual boxes.
Here we answer that question affirmatively for a broad class given by the so-called \emph{$b$-cycle boxes} 
(see Fig.\ \ref{figncycle}). A $b$-cycle box has as many maximal contexts as buttons ($b$), each $i$-th 
maximal context consists of two buttons ($i$ and $i+1$), each button belongs to two contexts 
($x_i=1$ for $\boldsymbol{x}^{(i)}$ and $\boldsymbol{x}^{(i-1)}$), and each $i$-th input button has two associated output lights, the $(2i-1)$-th and the $(2i)$-th lights, with $l=2b$. Modulo $b$ is implicitly assumed for the labels of buttons, contexts, and lights.
 These boxes admit $2^{b-1}$ contextuality bits:
 \begin{lem}[Existence of contextuality bits] 
 \label{lem:Cont_bit}
 For any $b\geq3$, all $b$-cycle boxes in $\mathsf{ND}$ can be freely obtained from a $b$-cycle box with behavior $\boldsymbol{P}^{(\boldsymbol{\gamma})}_{\mathcal{A}|\mathcal{X}}$ of components
 \begin{equation}
 \label{eq:ext_behav}
 p^{(\boldsymbol{\gamma})}_{\mathcal{A}|\mathcal{X}}(\boldsymbol{a},\boldsymbol{x})= 
\begin{cases}
 \frac{1}{2},& \text{if } x_{i}=1= x_{i+1} \text{ and }\\
 & a_{2i-\gamma_i}=a_{2(i+1)-\gamma_{i+1}},\\ 
 0, & \text{otherwise},
\end{cases}
\end{equation}
for all $i\in\mathcal{X}$, with $\boldsymbol{\gamma}\coloneqq(\gamma_1,\hdots, \gamma_b)$ such that $\gamma_i=0$ or 1 and $\|\boldsymbol{\gamma}\|_\mathrm{h}$ is an odd integer.
 \end{lem}
 
Eq.\ \eqref{eq:ext_behav} describes any of the $2^{b-1}$ contextual $b$-cycle behaviors extremal in $\mathsf{ND}$, derived (in a different notation) and shown to be equivalent under noncontextual relabelings of outputs in Ref.\ \cite{AQBTC13}. The proof of the lemma, given in App.\ \ref{sec:Ext_b_cycle}, consists then in showing that any convex combination of noncontextual relabelings of outputs can be carried out by a wiring in $\mathsf{NCW}$. 
For the particular case $b=4$ (the CHSH scenario), the behaviors in Eq.\ \eqref{eq:ext_behav} become equivalent to the no-signalling extremal Popescu-Rohrlich box \cite{PR94}, which is known to generate all other no-signalling boxes under local wirings assisted by shared randomness \cite{GWAN12,V14, GA17}. 
Lemma \ref{lem:Cont_bit} thus generalizes this fact to arbitrary $b\geq 3$ and noncontextual wirings.  
Finally, it is important to mention that, for even $b$, the buttons can be split into two disjoint subsets of $b/2$ incompatible buttons each, an the lights can be reduced from $2b$ to only 4 (one mutually exclusive pair per subset of buttons), as in the chained inequalities \cite{BC90}. This is an alternative representation of the same physical box. Our formalism is totally versatile in this sense, as it can directly deal with any chosen representation of a box. 

%%%%%%%%%%%%%%%%%%%%%%%%%%%%%%%%%%%%%%%%%%%%%%%%%%%%%%%%%%%%%%%%%%%%%

\emph{Final discussion}. 
%We have introduced the \lo{class} of noncontextual wirings. 
In contrast to more abstract approaches \cite{GHHHJKW14,HGJKL15}, noncontextual wirings admit a friendly analytical parametrization.
This is useful to classify, quantify, and operationally manipulate contextuality as a formal resource. For instance, monotonicity of contextuality under a non-trivial class of free operations was not clear for a long time. Here, we have solved this problem by showing that the relative entropy of contextuality is a contextuality monotone under noncontextual wirings. Furthermore, we have also shown that single, maximally-contextual boxes that serve as contextuality bits exist for all cycle boxes. Cycle boxes encompass important Bell scenarios \cite{CHSH69,KCBS08}; and the treatment can additionally be extended to bipartite boxes with more outputs \cite{Barrett05b}.
Interesting open questions are, e.g., what the simplest box admitting inequivalent (i.e, not freely interconvertible) classes of contextuality is and what the simplest one allowing for contextuality distillation.
Our findings provide the missing ingredient for a complete, unified resource theory of contextuality and Bell nonlocality.

%%%%%%%%%%%%%%%%%%%%%%%%%%%%%%%%%%%%%%%%%%%%%%%%%%%%%%%%%%%%%%%%%%%%%

\emph{Acknowledgements}. 
The present work was initiated during the workshop ``Quantum Correlations, Contextuality, and All That\ldots Again'' at the International 
Institute of Physics (IIP), Natal, Brazil. The participants of the workshop as well as the hospitality of IIP are gratefully acknowledged. 
We thank D. Cavalcanti, C. Duarte, and M. Pusey for fruitful discussions. BA thanks the Instituto de Matem\'atica Pura e Aplicada (IMPA) for the hospitality at Rio de Janeiro, Brazil. BA, MTC, and LA acknowledge financial support from the Brazilian ministries MEC and MCTIC and agencies CNPq, CAPES, FAPERJ, FAEPEX, and INCT-IQ. 
AC acknowledges support from Project No.\ FIS2014-60843-P, ``Advanced Quantum Information'' (MINECO, Spain), with FEDER funds,
the FQXi Large Grant ``The Observer Observed: A Bayesian Route to the Reconstruction of Quantum Theory,''
and the project ``Photonic Quantum Information'' (Knut and Alice Wallenberg Foundation, Sweden).

%
%%%%%%%%%%%%%%%%%%%%%%%%%%%%%%%%%%%%%%%%%%%%%%%%%%%%%%%%%%%%%%%%%%%%
%

\appendix

\section{Proof that the buttons in $\mathcal{X}_{[j]}\in\mathcal{X}$ and $\mathcal{Y}_{[j]}$ are all incompatible}
\label{sec:Append0}
Here, we explicitly prove that for all $j\in\mathcal{C}$, both $\mathcal{X}_{[j]}\in\mathcal{X}$ and $\mathcal{Y}_{[j]}\in\mathcal{Y}$, 
as defined right after Eq.\ \eqref{eq:def_post_box}, are composed exclusively of incompatible buttons according to
$\mathcal{I}_\mathcal{X}$ and $\mathcal{I}_\mathcal{Y}$. The proof is simple and consists of \emph{reductio ad absurdum}. 
Suppose that, for some $j\in\mathcal{C}$, not all buttons in $\mathcal{X}_{\left[ j\right]}$ are incompatible. This implies 
that the subset $\mathcal{A}_{\left[ j\right]}$ of lights in $\mathcal{A}$ associated with $j\in\mathcal{C}$ are not all mutually
exclusive. Since $\mathcal{A}_{\left[ j\right]}$ coincides with the subset of buttons $\mathcal{Z}_{\left( j\right)}$, that implies, 
in turn, that not all buttons in $\mathcal{Z}_{\left( j\right)}$ are incompatible. However, the latter is false by assumption
(because $\boldsymbol{B}_\mathrm{POST}\left(\boldsymbol{b}, \boldsymbol{y} \right)$ are well-defined boxes, so that no 
compatible buttons can share a common associated light). This proves that $\mathcal{X}_{[j]}\in\mathcal{X}$ contains 
always only incompatible buttons according to $\mathcal{I}_\mathcal{X}$. By the same reasoning, this implies that 
$\mathcal{Y}_{[j]}\in\mathcal{Y}$ contains always only incompatible according to $\mathcal{I}_\mathcal{Y}$, which finishes the proof.

\section{Proof of Lemma \ref{Lem:ND}}
\label{sec:Append1}

For any $\boldsymbol{y}, \boldsymbol{w}\in\mathcal{I}_\mathcal{Y}$ such that $\boldsymbol{y}\succeq\boldsymbol{w}$, consider the sum
\be 
\label{eq:Appen1_0}
\sum_{c_j; j \notin \mathcal{C}_{[\boldsymbol{w}]}}
p_{\mathcal{C}|\mathcal{Z}, \boldsymbol{b}, \boldsymbol{y} }\left(\boldsymbol{c},\boldsymbol{z}\right)
\ee
in which $\mathcal{C}_{[\boldsymbol{w}]}$ is the set of lights in $\mathcal{C}$ associated with the buttons in 
$\boldsymbol{w}$ for the resulting wired box $\mathcal{W}_\mathsf{NC}(\boldsymbol{B})$. Let $\tilde{\boldsymbol{a}}= \boldsymbol{a}_{(\boldsymbol{w})}$
and $\tilde{\boldsymbol{b}}= \boldsymbol{b}_{\left(\tilde{\boldsymbol{a}}\right)}$. 
For $j \in \mathcal{C}_{[\boldsymbol{w}]} $, we have that $\boldsymbol{a}_{(j)}= \tilde{\boldsymbol{a}_{(j)}}$,
$\boldsymbol{b}_{[j]}= \tilde{\boldsymbol{b}_{[j]}}$, and $\boldsymbol{y}_{[j]}= \tilde{\boldsymbol{y}_{[j]}}$,
otherwise we would have incompatible buttons in the same context, which is not possible. Hence, Eq.\ \eqref{eq:def_post_box} implies that 
\be \sum_{c_j; j \notin \mathcal{C}_{[\boldsymbol{w}]}}
p_{\mathcal{C}|\mathcal{Z}, \boldsymbol{b}, \boldsymbol{y} }(\boldsymbol{c},\boldsymbol{a})
= p_{\mathcal{C}|\mathcal{Z}, \tilde{\boldsymbol{b}}, \boldsymbol{w} }(\boldsymbol{c}_{[w]},\tilde{\boldsymbol{a}}). \ee
From this it follows that
\begin{widetext}
\begin{eqnarray}
\sum_{c_j; j \notin \mathcal{C}_{[\boldsymbol{w}]}} p_{\mathcal{C}|\mathcal{Y}}(\boldsymbol{c},\boldsymbol{y}) & = & 
\sum_{c_j; j \notin \mathcal{C}_{[\boldsymbol{w}]}} \sum_{\boldsymbol{\alpha},\boldsymbol{\beta}} p_{\mathcal{C}|\mathcal{Z}, \boldsymbol{b}, \boldsymbol{y} }(\boldsymbol{c},\boldsymbol{\alpha})\, p_{\mathcal{A}|\mathcal{X}}(\boldsymbol{\alpha},\boldsymbol{\beta})\, p_{\mathcal{B}|\mathcal{Y}}(\boldsymbol{\beta},\boldsymbol{y})\\ \nonumber
 & = & \sum_{\boldsymbol{\alpha},\boldsymbol{\beta}} 
p_{\mathcal{C}|\mathcal{Z}, \tilde{\boldsymbol{b}}, \boldsymbol{w} }(\boldsymbol{c}_{[w]},\tilde{\boldsymbol{\alpha}})
\, p_{\mathcal{A}|\mathcal{X}}(\boldsymbol{\alpha},\boldsymbol{\beta})\, p_{\mathcal{B}|\mathcal{Y}}(\boldsymbol{\beta},\boldsymbol{y}) \\\nonumber
 &= & \sum_{\tilde{\boldsymbol{\alpha}},\tilde{\boldsymbol{\beta}}} 
p_{\mathcal{C}|\mathcal{Z}, \tilde{\boldsymbol{\beta}}, \boldsymbol{y} }(\boldsymbol{c}_{[w]},\tilde{\boldsymbol{\alpha}})
\, p_{\mathcal{A}|\mathcal{X}}(\tilde{\boldsymbol{\alpha}},\tilde{\boldsymbol{\beta}})\, p_{\mathcal{B}|\mathcal{Y}}(\tilde{\boldsymbol{\beta}},\tilde{\boldsymbol{y}})\\ \nonumber
&=& p_{\mathcal{C}_{[w]}|\mathcal{Y}}(\boldsymbol{c}_{[w]}, \boldsymbol{w}).
\end{eqnarray}
\end{widetext}
 This concludes the proof that the resulting wired box $\mathcal{W}_\mathsf{NC}(\boldsymbol{B})$ is nondisturbing. 

%%%%%%%%%%%%%%%%%%%%%%%%%%%%%%%%%%%%%%%%%%%%%%%%%%%%%%%%%%%%%%%%%%%

\section{Proof of Theorem 2}
\label{sec:app_2}

% The assumptions over the post-processing imply the existence of variables $\lambda_\mathrm{POST}$ such that 
% 
% 
% \begin{align} p_\mathrm{POST}\left( \boldsymbol{z}|\boldsymbol{b},\boldsymbol{a},\boldsymbol{x}\right)&= &\sum_{\lambda_\mathrm{POST}} p\left(\lambda_\mathrm{POST}\right) S\left(\boldsymbol{z}|\boldsymbol{b},\boldsymbol{a},\boldsymbol{x},\lambda_\mathrm{POST}\right) \nonumber \\
% &= &\sum_{\lambda_\mathrm{POST}} p\left(\lambda_\mathrm{POST}\right) \prod_{z_j \in \boldsymbol{z}} D_j\left(z_j|\boldsymbol{b},\boldsymbol{a},\boldsymbol{x},\lambda_\mathrm{POST}\right)
% \end{align}

Assuming Eqs.\ \eqref{eq:def_non_context_ala_Bell} and \eqref{eq:def_post_box}, and the analogous equation for $\boldsymbol{B}_\mathrm{PRE}$: for all $\boldsymbol{b}\in\{0,1\}^{\mathcal{B}}$ and $\boldsymbol{y}\in\mathcal{I}_\mathcal{Y}$,
\begin{eqnarray}
p_{\mathcal{B}|\mathcal{Y}}(\boldsymbol{b},\boldsymbol{y})=
%\nonumber
%\sum_{\gamma} p\left(\gamma\right)\, D (\boldsymbol{b}\vert\boldsymbol{y},\lambda)
%\coloneqq&\\
\sum_{\gamma} p_\Gamma(\gamma)\, \prod_{j\in\mathcal{B}}\, D_j (b_j\vert \boldsymbol{y}_{(j)},\gamma),
\end{eqnarray} 
we need to prove that the final behavior in Eq.\ \eqref{eqwiring} is noncontextual.

To this end, let us first introduce the short-hand notation $\boldsymbol{\lambda}\coloneqq(\gamma,\lambda,\phi)$ and 
$p_{\boldsymbol{\Lambda}}(\boldsymbol{\lambda})\coloneqq p_{\Gamma}(\gamma)\, p_{\Lambda}(\lambda)\,p_{\Phi}(\phi)$. Then, we explicitly write out Eq.\ \eqref{eqwiring} as
\begin{widetext}
\begin{eqnarray}
\label{eq:wirings_DCHV}
\nonumber
p_{\mathcal{C}|\mathcal{Y}}(\boldsymbol{c},\boldsymbol{y}) &=&
\sum_{\substack{\boldsymbol{\alpha}\in\{0,1\}^{|\mathcal{A}|} \\ \boldsymbol{\beta}\in\{0,1\}^{|\mathcal{B}|}}} 
 p_{\mathcal{C}|\mathcal{Z}, \boldsymbol{\beta}, \boldsymbol{y} }(\boldsymbol{c},
\boldsymbol{\alpha}) \, p_{\mathcal{A}|\mathcal{X}}(\boldsymbol{\alpha},\boldsymbol{\beta})\, p_{\mathcal{B}|\mathcal{Y}}(\boldsymbol{\beta},\boldsymbol{y})\\
\nonumber
&=&\sum_{\substack{\boldsymbol{\alpha}\in\{0,1\}^{|\mathcal{A}|} \\ \boldsymbol{\beta}\in\{0,1\}^{|\mathcal{B}|}}}\sum_{\boldsymbol{\lambda}} p_{\boldsymbol{\Lambda}}(\boldsymbol{\lambda}) \,
\prod_{j\in\mathcal{C}}\, D_j (c_j\vert \boldsymbol{\alpha}_{(j)}, \boldsymbol{\beta}_{[j]}, \boldsymbol{y}_{[j]},\phi)
\prod_{k\in\mathcal{A}}\, D_k (\alpha_k\vert \boldsymbol{\beta}_{(k)},\lambda)\,
\prod_{i\in\mathcal{B}}\, D_i (\beta_i\vert \boldsymbol{y}_{(i)},\gamma)\,
\\
%&=&\sum_{\boldsymbol{\lambda}} p_{\boldsymbol{\Lambda}}(\boldsymbol{\lambda})\,
%D (\boldsymbol{c}\vert \boldsymbol{\lambda}, \tilde{\boldsymbol{a}},\tilde{\boldsymbol{b}},\boldsymbol{y})\\
&=& \sum_{\boldsymbol{\lambda}} p_{\boldsymbol{\Lambda}}(\boldsymbol{\lambda})\,
\prod_{j\in\mathcal{C}} D_j \left(c_j\Big\vert \tilde{\boldsymbol{\alpha}}_{(j)}\left(\tilde{\boldsymbol{\beta}}_{[j]}(\boldsymbol{y}_{[j]},\gamma),\lambda\right),\tilde{\boldsymbol{\beta}}_{[j]}(\boldsymbol{y}_{[j]},\gamma),\boldsymbol{y}_{[j]},\phi\right),
\end{eqnarray}
\end{widetext}
where the sums over $\boldsymbol{a}$ and $\boldsymbol{b}$ disappear because of the deterministic response functions.
Besides, $\tilde{\boldsymbol{\beta}}_{[j]}(\boldsymbol{y}_{[j]},\gamma)$ is the deterministic output substring of the pre-processing box as a function of the input substring $\boldsymbol{y}$ and the deterministic strategy $\gamma$, and $\tilde{\boldsymbol{\alpha}}_{(j)}\left(\tilde{\boldsymbol{\beta}}_{[j]}(\boldsymbol{y}_{[j]},\gamma),\lambda\right)$ is the deterministic output substring of the initial box as a function of the input substring $\tilde{\boldsymbol{\beta}}_{[j]}(\boldsymbol{y}_{[j]},\gamma)$ and the deterministic strategy $\lambda$.

Identifying the $\boldsymbol{\lambda}$-th NC deterministic response function for the output light 
$j\in\mathcal{C}$ given the input substring of buttons in $\mathcal{Y}$ associated with light $j\in\mathcal{C}$ as 
$D_j (c_j\vert \boldsymbol{y}_{[j]},\boldsymbol{\lambda})\coloneqq D_j \left(c_j\Big\vert \tilde{\boldsymbol{\alpha}}_{(j)}\left(\tilde{\boldsymbol{\beta}}_{[j]}(\boldsymbol{y}_{[j]},\gamma),\lambda\right),\tilde{\boldsymbol{\beta}}_{[j]}(\boldsymbol{y}_{[j]},\gamma),\boldsymbol{y}_{[j]},\phi\right)$, we write Eq.\ \eqref{eq:wirings_DCHV} as
\begin{align}
\label{eq:wirings_DCHV_final}
\nonumber
p_{\mathcal{C}|\mathcal{Y}}(\boldsymbol{c},\boldsymbol{y})
&=\sum_{\boldsymbol{\lambda}} p_{\boldsymbol{\Lambda}}(\boldsymbol{\lambda}) D (\boldsymbol{c}\vert\boldsymbol{\lambda}, \boldsymbol{y})\\
 &= \sum_{\boldsymbol{\lambda}} p_{\boldsymbol{\Lambda}}(\boldsymbol{\lambda}) \prod_{j\in\mathcal{C}} D_j (c_j\vert \boldsymbol{y}_{[j]},\boldsymbol{\lambda}),
\end{align}
which is manifestly in the explicit form of a NC hidden-variable model. This concludes the proof.

%\lo{
%Apart from an extensive check-up of all my calculations, there are a number of things that we (or at least I) must still understand better:
%\begin{itemize}
%\item The issue about the characterization of the output exclusivity hyper-graph of the final composed box. I mean, is it really $\mathcal{O}_\mathcal{C}\subseteq\mathcal{O}^{(3)}_\mathcal{C}$ or $\mathcal{O}_\mathcal{C}=\mathcal{O}^{(3)}_\mathcal{C}$? This is, in turn, probably, very connected to the first item.
%\item And this is also connected to the previous two items: If it is in general $\mathcal{O}_\mathcal{C}\subseteq\mathcal{O}^{(3)}_\mathcal{C}$, how do we then determine $\mathcal{O}_\mathcal{C}$ when the initial box is contextual? We should at least be able to solve this issue for the most relevant case of the initial box being a generic nondisturbing box. Note that, for noncontextual initial boxes, as done above, this is easy because these boxes are always a convex combination of the deterministic extremal points of the nondisturbance polytope. However, generic nondisturbing boxes involve convex combinations of all extremal points of the nondisturbance polytope, the deterministic noncontextual ones and the nondeterministic contextual ones. Anyway, I think it should become clear as we think about it a bit :-)
%\end{itemize}
%}

%%%%%%%%%%%%%%%%%%%%%%%%%%%%%%%%%%%%%%%%%%%%%%%%%%%%%%%%%%%%%%%%%%%

\section{Proof of Lemma \ref{Lem:monotonicity}}
\label{sec:monotonicity}
We begin with the definition of $S\left(\boldsymbol{B}\| \boldsymbol{B}^{*}\right)$ appearing in Eq.\ \eqref{eqcontextmonotone}.
The relative entropy, or Kullback-Leibler divergence, $S_\mathrm{d}$ of a probability distribution $\boldsymbol{P}$ (over a sample space $\Omega$) relative to another probability distribution $\boldsymbol{P}^*$ (over the same sample space) is defined as
 \be
 \label{eq:rel_ent_dist_def}
 S_\mathrm{d}(\boldsymbol{P}\|\boldsymbol{P}^*) = \sum_{i\in \Omega} p(i) \, \log\left(\frac{p(i)}{p^*(i)}\right).
 \ee 
With this, one can define the relative entropy $S_\mathrm{b}$ between two behaviors $\boldsymbol{P}_{\mathcal{A}|\mathcal{X}}$ and $\boldsymbol{P}_{\mathcal{A}|\mathcal{X}}^*$ as the relative entropy $S_\mathrm{d}$ between the output distributions obtained from $\boldsymbol{P}_{\mathcal{A}|\mathcal{X}}$ and $\boldsymbol{P}_{\mathcal{A}|\mathcal{X}}^*$ for the optimal input choice \cite{GA17}:
\begin{align}
\label{eq:rel_ent_beha_def}
S_{\rm{b}}\left(\boldsymbol{P}_{\mathcal{A}|\mathcal{X}}\|\boldsymbol{P}_{\mathcal{A}|\mathcal{X}}^*\right)\coloneqq\max_{\boldsymbol{x}\in\mathcal{I}_\mathcal{X}} S_{{\rm{d}}}\left({\boldsymbol{P}_{\mathcal{A}|\mathcal{X}}}(\cdot, \boldsymbol{x})\big\|{\boldsymbol{P}*_{\mathcal{A}|\mathcal{X}}}(\cdot, \boldsymbol{x})\right).
\end{align}
In turn, the (box) relative entropy $S$ between two nondisturbing boxes $\boldsymbol{B}\coloneqq\{\mathcal{I}_\mathcal{X},\mathcal{O}_\mathcal{A}, %\mathcal{X}_{\mathcal{A}},
\boldsymbol{P}_{\mathcal{A}|\mathcal{X}}\}$ and $\boldsymbol{B}^{*}\coloneqq\{\mathcal{I}_\mathcal{X},\mathcal{O}_\mathcal{A}, % \mathcal{X}_{\mathcal{A}}, 
\boldsymbol{P}_{\mathcal{A}|\mathcal{X}}^*\}$ is defined as the (behavior) relative entropy between their respective behaviors \cite{GHHHJKW14}: 
\begin{align}
\label{eq:rel_ent_box_def}
S\left(\boldsymbol{B}\| \boldsymbol{B}^{*}\right)\coloneqq S_{\rm{b}}\left(\boldsymbol{P}_{\mathcal{A}|\mathcal{X}}\|\boldsymbol{P}_{\mathcal{A}|\mathcal{X}}^*\right).
\end{align}

Now we proceed to prove the lemma. First we show monotonicity of the box relative entropy $S$ under an arbitrary noncontextual wiring $\mathcal{W}_\mathsf{NC} \in \mathsf{NC}$ . 
Given $\boldsymbol{B}\coloneqq\{\mathcal{I}_\mathcal{X},\mathcal{O}_\mathcal{A}, \boldsymbol{P}_{\mathcal{A}|\mathcal{X}}\}\in\mathsf{NC}$
and $\boldsymbol{B}'\coloneqq\{\mathcal{I}_\mathcal{X},\mathcal{O}_\mathcal{A}, \boldsymbol{P}'_{\mathcal{A}|\mathcal{X}}\}\in\mathsf{NC}$,
let $\boldsymbol{B}_\mathrm{f}\coloneqq\mathcal{W}_\mathsf{NC}(\boldsymbol{B})=\{\mathcal{I}_\mathcal{Y},\mathcal{O}_\mathcal{C}, \boldsymbol{P}_{\mathcal{C}|\mathcal{Y}}\}$ and %\mathcal{Y}_{\mathcal{C}},
$\boldsymbol{B}'_\mathrm{f}\coloneqq\mathcal{W}_\mathsf{NC}(\boldsymbol{B}')=\{\mathcal{I}_{\mathcal{Y}},\mathcal{O}_{\mathcal{C}}, 
\boldsymbol{P}'_{\mathcal{C}|\mathcal{Y}}\}$. In addition, denote by $\boldsymbol{y}^*$ the string in $\mathcal{I}_\mathcal{Y}$ such that 
\begin{align}
\nonumber
&S_\mathrm{d}\left(\boldsymbol{P}_{\mathcal{C}|\mathcal{Y}} (\cdot, \boldsymbol{y}^*) \| \boldsymbol{P}'_{\mathcal{C}|\mathcal{Y}} (\cdot, \boldsymbol{y}^*)\right)\coloneqq\\
\label{eq:auxiliary_y}
&\max_{\boldsymbol{y}\in\mathcal{I}_\mathcal{Y}} S_\mathrm{d}\left(\boldsymbol{P}_{\mathcal{C}|\mathcal{Y}} (\cdot, \boldsymbol{y}) \| \boldsymbol{P}'_{\mathcal{C}|\mathcal{Y}} (\cdot, \boldsymbol{y})\right). 
\end{align}
Then,
\begin{widetext}
\begin{align}
 S\left(\boldsymbol{B}_\mathrm{f} \left\| \boldsymbol{B}'_\mathrm{f} \right. \right) = & 
 \max_{\boldsymbol{y}{\in\mathcal{I}_\mathcal{Y}}} S_\mathrm{d}\left({\boldsymbol{P}}_{\mathcal{C}|\mathcal{Y}} (\cdot, {\boldsymbol{y}}) \left\| {\boldsymbol{P}'}_{\mathcal{C}|\mathcal{Y}} (\cdot, {\boldsymbol{y}})\right.\right\} &\label{eq:def_S}\\
 =& \sum_{\boldsymbol{c}} p_{\mathcal{C}|\mathcal{Y}} ({\boldsymbol{c},\boldsymbol{y}^*})\log \left( \frac{p_{\mathcal{C}|\mathcal{Y}} ({\boldsymbol{c},\boldsymbol{y}^*})}{p'_{\mathcal{C}|\mathcal{Y}} ({\boldsymbol{c},\boldsymbol{y}^*})}\right)& \label{eq:def_S_d}\\
=& \sum_{\substack{\boldsymbol{\alpha}, \boldsymbol{\beta}, \boldsymbol{c}}} p_{\mathcal{C}|\mathcal{Z}, \boldsymbol{\beta}, \boldsymbol{y}^* }(\boldsymbol{c},\boldsymbol{\alpha}) \, p_{\mathcal{A}|\mathcal{X}}(\boldsymbol{\alpha},\boldsymbol{\beta})\, p_{\mathcal{B}|\mathcal{Y}}(\boldsymbol{\beta},\boldsymbol{y}^*)
\log \left( \frac{\displaystyle{\sum_{\substack{\boldsymbol{\alpha}, \boldsymbol{\beta}}}} p_{\mathcal{C}|\mathcal{Z}, \boldsymbol{\beta}, \boldsymbol{y}^* }(\boldsymbol{c},\boldsymbol{\alpha}) \, p_{\mathcal{A}|\mathcal{X}}(\boldsymbol{\alpha},\boldsymbol{\beta})\, p_{\mathcal{B}|\mathcal{Y}}(\boldsymbol{\beta},\boldsymbol{y}^*)}{\displaystyle{\sum_{\substack{\boldsymbol{\alpha}, \boldsymbol{\beta}}}} p_{\mathcal{C}|\mathcal{Z}, \boldsymbol{\beta}, \boldsymbol{y}^* }(\boldsymbol{c},\boldsymbol{\alpha}) \, p'_{\mathcal{A}|\mathcal{X}}(\boldsymbol{\alpha},\boldsymbol{\beta})\, p_{\mathcal{B}|\mathcal{Y}}(\boldsymbol{\beta},\boldsymbol{y}^*)}\right)&\label{eq:dists_sum}\\
\leq & \sum_{\substack{\boldsymbol{\alpha}, \boldsymbol{\beta},\boldsymbol{c}}} p_{\mathcal{C}|\mathcal{Z}, \boldsymbol{\beta}, \boldsymbol{y}^* }(\boldsymbol{c},\boldsymbol{\alpha}) \, p_{\mathcal{A}|\mathcal{X}}(\boldsymbol{\alpha},\boldsymbol{\beta})\, p_{\mathcal{B}|\mathcal{Y}}(\boldsymbol{\beta},\boldsymbol{y}^*)
\log \left( \frac{ p_{\mathcal{C}|\mathcal{Z}, \boldsymbol{\beta}, \boldsymbol{y}^* }(\boldsymbol{c},\boldsymbol{\alpha}) \, p_{\mathcal{A}|\mathcal{X}}(\boldsymbol{\alpha},\boldsymbol{\beta})\, p_{\mathcal{B}|\mathcal{Y}}(\boldsymbol{\beta},\boldsymbol{y}^*)}{ p_{\mathcal{C}|\mathcal{Z}, \boldsymbol{\beta}, \boldsymbol{y}^* }(\boldsymbol{c},\boldsymbol{\alpha}) \, p'_{\mathcal{A}|\mathcal{X}}(\boldsymbol{\alpha},\boldsymbol{\beta})\, p_{\mathcal{B}|\mathcal{Y}}(\boldsymbol{\beta},\boldsymbol{y}^*)}\right)&\label{eq:prop_sum}\\
=& \sum_{\substack{\boldsymbol{\alpha}, \boldsymbol{\beta}, \boldsymbol{c}}} p_{\mathcal{C}|\mathcal{Z}, \boldsymbol{\beta}, \boldsymbol{y}^* }(\boldsymbol{c},\boldsymbol{\alpha}) \, p_{\mathcal{A}|\mathcal{X}}(\boldsymbol{\alpha},\boldsymbol{\beta})\, p_{\mathcal{B}|\mathcal{Y}}(\boldsymbol{\beta},\boldsymbol{y}^*)
\log \left( \frac{ p_{\mathcal{A}|\mathcal{X}}(\boldsymbol{\alpha},\boldsymbol{\beta})}{ p'_{\mathcal{A}|\mathcal{X}}(\boldsymbol{\alpha},\boldsymbol{\beta})}\right)&\label{eq:cancel}\\
=& \sum_{\substack{\boldsymbol{\alpha}, \boldsymbol{\beta}}} p_{\mathcal{A}|\mathcal{X}}(\boldsymbol{\alpha},\boldsymbol{\beta})\, p_{\mathcal{B}|\mathcal{Y}}(\boldsymbol{\beta},\boldsymbol{y}^*)
\log \left( \frac{ p_{\mathcal{A}|\mathcal{X}}(\boldsymbol{\alpha},\boldsymbol{\beta})}{ p'_{\mathcal{A}|\mathcal{X}}(\boldsymbol{\alpha},\boldsymbol{\beta})}\right) &\label{eq:sum_prob} \\
=& \sum_{ \boldsymbol{\beta}} p_{\mathcal{B}|\mathcal{Y}}(\boldsymbol{\beta},\boldsymbol{y}^*) \sum_{\substack{\boldsymbol{\alpha}}}
 p_{\mathcal{A}|\mathcal{X}}(\boldsymbol{\alpha},\boldsymbol{\beta})\, 
\log \left( \frac{ p_{\mathcal{A}|\mathcal{X}}(\boldsymbol{\alpha},\boldsymbol{\beta})}{ p'_{\mathcal{A}|\mathcal{X}}(\boldsymbol{\alpha},\boldsymbol{\beta})}\right)&\label{eq:split_ab}\\
=& \sum_{ \boldsymbol{\beta}} p_{\mathcal{B}|\mathcal{Y}}(\boldsymbol{\beta},\boldsymbol{y}^*) \,
S_\mathrm{d}\left({\boldsymbol{P}}_{\mathcal{A}|\mathcal{X}} (\cdot, {\boldsymbol{\beta}}) \left\| {\boldsymbol{P}'}_{\mathcal{A}|\mathcal{X}} (\cdot, {\boldsymbol{\beta}})\right.\right)& \label{eq:def_S_d_2}\\
\leq & \max_{\boldsymbol{x}{\in\mathcal{I}_\mathcal{X}} } 
S_\mathrm{d}\left({\boldsymbol{P}}_{\mathcal{A}|\mathcal{X}} (\cdot, {\boldsymbol{x}}) \left\| {\boldsymbol{P}'}_{\mathcal{A}|\mathcal{X}} (\cdot, {\boldsymbol{x}})\right.\right) &\label{eq:max<mean}\\
%&= &\max_{\boldsymbol{x}\lo{\in\mathcal{I}_\mathcal{X}} } 
%S_\mathrm{d}\left(\lo{\boldsymbol{P}}_{\mathcal{A}|\mathcal{X}} (\cdot, \lo{\boldsymbol{x}}) \left\| \lo{\boldsymbol{P}'}_{\mathcal{A}|\mathcal{X}} (\cdot, \lo{\boldsymbol{x}})\right.\right) \label{eq:max_y}\\
=& S\left(\boldsymbol{B} \left\| \boldsymbol{B}' \right. \right).\label{eq:def_S_d_3}&\end{align}
\end{widetext}
Eqs.\ \eqref{eq:def_S} follows from the definition of $S$, 
Eq.\ \eqref{eq:def_S_d} from Eq.\ \eqref{eq:auxiliary_y}, Eq.\ \eqref{eq:dists_sum} from Eq.\ \eqref{eqwiring}, Eq.\ \eqref{eq:prop_sum} from the log sum inequality $\sum_i x_i \log \left(\frac{\sum_i x_i}{\sum_i y_i}\right) \leq 
\sum_i x_i \log \left(\frac{ x_i}{y_i}\right)$, Eqs.\ \eqref{eq:cancel} and \eqref{eq:split_ab} from basic algebra, Eq.\ \eqref{eq:sum_prob} from summing over $\boldsymbol{c}$ and the 
fact that $p_{\mathcal{C}|\mathcal{Z}, \boldsymbol{\beta}, \boldsymbol{y} }(\cdot,\boldsymbol{\alpha})$ is a well-normalized probability distribution,
Eqs.\ \eqref{eq:def_S_d_2} from the definition of $S_\mathrm{b}$,
Eq.\ \eqref{eq:max<mean} from the fact the average is smaller than the largest value,
and Eq.\ \eqref{eq:def_S_d_3} from the definition of $S_\mathrm{d}$.

Now we can prove monotonicity of $R_C$. Let $\boldsymbol{B}' \in \mathsf{NC}$ be the noncontextual box for wich the minumum in Eq.\ \eqref{eqcontextmonotone} is
achieved for box $\boldsymbol{B}$, that is,
such that $R_C \left(\boldsymbol{B}\right) = S\left(\boldsymbol{B} \left\| \boldsymbol{B}'\right. \right)$. Then, we have
\begin{align}
 R_C \left(\boldsymbol{B}_\mathrm{f}\right) = & \min_{\boldsymbol{B}^* \in NC} S\left(\boldsymbol{B}_\mathrm{f} \left\| \boldsymbol{B}* \right. \right)& \\
 \leq & \ S\left(\boldsymbol{B}_\mathrm{f} \left\| \boldsymbol{B}'_\mathrm{f} \right. \right) & \\
 \leq & \ S\left(\boldsymbol{B} \left\| \boldsymbol{B}'\right. \right) &\\
 = & \ R_C \left(\boldsymbol{B}\right),&
\end{align}
which concludes the proof.

%%%%%%%%%%%%%%%%%%%%%%%%%%%%%%%%%%%%%%%%%%%%%%%%%%%%%%%%%%%%%%%%%%%

\section{Proof of Lemma \ref{lem:Cont_bit}}
\label{sec:Ext_b_cycle}

Eq.\ \eqref{eq:ext_behav} is the expression, in our notation, of the $2^{b-1}$ extremal contextual behaviors of the $b$-cycle scenario derived in Theorem 2 of Ref.\ \cite{AQBTC13}. 
In turn, the $2^b$ extremal noncontextual behaviors
$\boldsymbol{P}^{(\boldsymbol{\zeta})}_{\mathcal{A}|\mathcal{X}}$ are expressed \cite{AQBTC13} as
\begin{equation}
 \label{eq:ext_nc_behav}
 p^{(\boldsymbol{\zeta})}_{\mathcal{A}|\mathcal{X}}(\boldsymbol{a},\boldsymbol{x})= 
\begin{cases}
 1,& \text{if } x_{i}=1 \text{ and } a_{2i-1}= \zeta_{i} + 1, \\
 0, & \text{otherwise},
\end{cases}
\end{equation}
where $\boldsymbol{\zeta} = \left( \zeta_1,\ldots , \zeta_{b}\right)$ is an arbitrary bit string of length $b$ that encodes the deterministic
output of each button $i$. Note also that $a_{2i-1}= \zeta_{i} + 1$ is equivalent to $a_{2i}= \zeta_{i}$.
The $2^{b-1}$ behaviors given by Eq.\ \eqref{eq:ext_behav}, together with the $2^{b}$ behaviors of Eq.\ \eqref{eq:ext_nc_behav}, constitute all the extremal points of $\mathsf{ND}$ \cite{AQBTC13}. Hence, any nondisturbing $b$-cycle behavior is a convex combination of $\boldsymbol{P}^{(\boldsymbol{\zeta})}_{\mathcal{A}|\mathcal{X}}$'s and $\boldsymbol{P}^{(\boldsymbol{\gamma})}_{\mathcal{A}|\mathcal{X}}$'s. Thus, we must prove that any particular $\boldsymbol{P}^{(\boldsymbol{\gamma'})}_{\mathcal{A}|\mathcal{X}}$ can be mapped by a wiring in $\mathsf{NCW}$ to an arbitrary convex combination of $\boldsymbol{P}^{(\boldsymbol{\zeta})}_{\mathcal{A}|\mathcal{X}}$'s and $\boldsymbol{P}^{(\boldsymbol{\gamma})}_{\mathcal{A}|\mathcal{X}}$'s.

% $\boldsymbol{P}^{(\boldsymbol{\gamma}')}_{\mathcal{A}|\mathcal{X}}$ and 
% $\boldsymbol{P}^{(\boldsymbol{\zeta}')}_{\mathcal{A}|\mathcal{X}}$, for arbitrary 
% 
% .

First we show that every behavior $\boldsymbol{P}^{(\boldsymbol{\gamma})}_{\mathcal{A}|\mathcal{X}}$ or 
$\boldsymbol{P}^{(\boldsymbol{\zeta})}_{\mathcal{A}|\mathcal{X}}$ can be obtained from any fixed $\boldsymbol{P}^{(\boldsymbol{\gamma'})}_{\mathcal{A}|\mathcal{X}}$
using a wiring in $\mathsf{NCW}$, for arbitrary $\boldsymbol{\gamma}$ and $\boldsymbol{\zeta}$.
For $\boldsymbol{P}^{({\boldsymbol{\gamma}})}_{\mathcal{A}|\mathcal{X}}$ we use as pre-processing the trivial (identity) deterministic box with 
$b$ input buttons and $b$ output lights, where each input $i$ has only one possible output $i$, and as post-processing the 
deterministic box
$
\boldsymbol{B}_{POST}^{\boldsymbol{\gamma'} \rightarrow \boldsymbol{\gamma}}$ with $2b$ input buttons and $2b$ output lights 
that permutes each $i$-th pair of lights, $2i-1$ and $2i$, whenever $\boldsymbol{\gamma}_i \neq \boldsymbol{\gamma'}_i$.
For $\boldsymbol{P}^{(\boldsymbol{\zeta})}_{\mathcal{A}|\mathcal{X}}$ we also use as pre-processing the trivial identity box and as post-processing the 
deterministic box
$
\boldsymbol{B}_{POST}^{\boldsymbol{\gamma'} \rightarrow \boldsymbol{\zeta}}$ with $2b$ input buttons and $2b$ output lights 
such that for each $i$-th pair of input (incompatible) buttons, $2i-1$ and $2i$, has light $2i-1+\boldsymbol{\zeta}_i$ as deterministic output.

Finally, taking the trivial box as pre-processing and an arbitrary convex combination of the post-processing boxes $
\boldsymbol{B}_{POST}^{\boldsymbol{\gamma'} \rightarrow \boldsymbol{\gamma}}$ and
$
\boldsymbol{B}_{POST}^{\boldsymbol{\gamma'} \rightarrow \boldsymbol{\zeta}}$ described above defines a class of wirings in $\mathsf{NCW}$
that takes $\boldsymbol{P}^{(\boldsymbol{\gamma'})}_{\mathcal{A}|\mathcal{X}}$ to all convex combinations of the $\boldsymbol{P}^{(\boldsymbol{\gamma})}_{\mathcal{A}|\mathcal{X}}$'s and 
$\boldsymbol{P}^{(\boldsymbol{\zeta})}_{\mathcal{A}|\mathcal{X}}$'s, as desired.


\begin{thebibliography}{99}

\bibitem{Specker60}
E. P. Specker,
{\it Die Logik nicht gleichzeitig entscheidbarer Aussagen},
\href{http://onlinelibrary.wiley.com/doi/10.1111/j.1746-8361.1960.tb00422.x/abstract}{Dialectica \textbf{14}, 239 (1960)}
[The logic of non-simultaneously decidable propositions, \href{http://arxiv.org/abs/1103.4537}{\eprint{arXiv:1103.4537}].}

\bibitem{KS67}
S. Kochen and E. P. Specker,
{\it The problem of hidden variables in quantum mechanics},
J. Math. Mech. \textbf{17}, 59 (1967).

\bibitem{Bell66}
J. S. Bell,
{\it On the problem of hidden variables in quantum mechanics},
\href{http://dx.doi.org/10.1103/RevModPhys.38.447}{Rev. Mod. Phys. \textbf{38}, 447 (1966).}

\bibitem{HLBBR06}
Y. Hasegawa, R. Loidl, G. Badurek, M. Baron, and H. Rauch, 
{\it Quantum Contextuality in a Single-Neutron Optical Experiment},
\href{http://dx.doi.org/10.1103/PhysRevLett.97.230401}{Phys. Rev. Lett. \textbf{97}, 230401 (2006).}

\bibitem{KZGKGCBR09}
G. Kirchmair, F. Z\"ahringer, R. Gerritsma, M. Kleinmann, O. G{\"u}hne, A. Cabello, R. Blatt, and C. F. Roos,
{\it State-independent experimental test of quantum contextuality},
\href{http://doi:10.1038/nature08172}{Nature (London) \textbf{460}, 494 (2009)}.

% \bibitem{BKSSCRH09}
% H. Bartosik, J. Klepp, C. Schmitzer, S. Sponar, A. Cabello, H. Rauch Y. and Hasegawa,
% Experimental Test of Quantum Contextuality in Neutron Interferometry,
% \href{10.1103/PhysRevLett.103.040403}{Phys. Rev. Lett. \textbf{103}, 040403 (2009)}.

\bibitem{ARBC09}
E. Amselem, M. R{\aa }dmark, M. Bourennane, and A. Cabello,
{\it State-Independent Quantum Contextuality with Single Photons},
\href{http://dx.doi.org/10.1103/PhysRevLett.103.160405}{Phys. Rev. Lett. \textbf{103}, 160405 (2009).}

\bibitem{LLSLRWZ11}
R. {\L}apkiewicz, P. Li, C. Schaeff, N. Langford, S. Ramelow, M. Wie\'sniak, and A. Zeilinger,
{\it Experimental non-classicality of an indivisible quantum system},
\href{http://dx.doi.org/doi:10.1038/nature10119}{Nature (London) \textbf{474}, 490 (2011).}

\bibitem{BCAFACTP13}
G. Borges, M. Carvalho, P.-L. de Assis, J. Ferraz, M. Ara\'ujo, A. Cabello, M. Terra Cunha, and S. P\'adua,
{\it Quantum contextuality in a Young-type interference experiment},
\href{https://doi.org/10.1103/PhysRevA.89.052106}{Phys. Rev. A \textbf{89}, 052106 (2014).}

\bibitem{HWVE14}
M. Howard, J. J. Wallman, V. Veitch, and J. Emerson,
{\it Contextuality supplies the `magic' for quantum computation},
\href{http://dx.doi.org/10.1038/nature13460}{Nature (London) \textbf{510}, 351 (2014).}

%\bibitem{CACBGXLC16}
%G. Ca\~nas, E. Acu\~na, J. Cari\~ne, J. F. Barra, E. S. G\'omez, G.B. Xavier, G. Lima and A. Cabello,
%{\it Experimental demonstration of the connection between quantum contextuality and graph theory},
%\href{10.1103/PhysRevA.94.012337}{Phys. Rev. A 94, 012337 (2016).} \la{Can we remove this reference before sending it to Adan? It is neither an important reference in general nor particularly connected to what we do here.}

%\bibitem{Ekert91}
%A. K. Ekert,
%Quantum cryptography based on Bell's theorem,
%\href{http://dx.doi.org/10.1103/PhysRevLett.67.661}{Phys. Rev. Lett. \textbf{67}, 661 (1991).}
%
%\bibitem{BHK05}
%J. Barrett, L. Hardy, and A. Kent,
%No signaling and quantum key distribution,
%\href{http://dx.doi.org/10.1103/PhysRevLett.95.010503}{Phys. Rev. Lett. \textbf{95}, 010503 (2005).}

\bibitem{Raussendorf13}
R. Raussendorf,
{\it Contextuality in measurement-based quantum computation},
\href{http://dx.doi.org/10.1103/PhysRevA.88.022322}{Phys. Rev. A \textbf{88}, 022322 (2013)}.

\bibitem{DGBR14}
N. Delfosse, P. A. Guerin, J. Bian, and R. Raussendorf,
{\it Wigner function negativity and contextuality in quantum computation on rebits},
\href{http://dx.doi.org/10.1103/PhysRevX.5.021003}{Phys. Rev. X \textbf{5}, 021003 (2015).}

\bibitem{UZZWYDDK13}
M. Um, Mark, X. Zhang, J. Zhang, Y. Wang, S. Yangchao, D. Deng, L. Duan, and K. Kim, 
{\it Experimental certification of random numbers via quantum contextuality},
\href{http://dx.doi.org/10.1038/srep01627}{Sci. Rep. \textbf{3}, 1627 (2013).}

\bibitem{Brunner13} 
N. Brunner, D. Cavalcanti, S. Pironio, V. Scarani, and S. Wehner, 
{\it Bell nonlocality}, 
\href{https://doi.org/10.1103/RevModPhys.86.419}{Rev. Mod. Phys. \textbf{86}, 419 (2014).}

\bibitem{GHHHJKW14}
A. Grudka, K. Horodecki, M. Horodecki, P. Horodecki, R. Horodecki, P. Joshi, W. K{\l}obus, and A. W\'ojcik,
{\it Quantifying Contextuality},
\href{http://dx.doi.org/10.1103/PhysRevLett.112.120401}{Phys. Rev. Lett. \textbf{112}, 120401 (2014).}

\bibitem{HGJKL15}
K. Horodecki, A. Grudka, P. Joshi, W. K{\l}obus, and J. {\L}odyga,
{\it Axiomatic approach to contextuality and nonlocality},
\href{http://dx.doi.org/10.1103/PhysRevA.92.032104}{Phys. Rev. A \textbf{92}, 032104 (2015).}

\bibitem{GWAN12}
R. Gallego, L. E. W\"{u}rflinger, A. Ac\'{i}n, and M. Navascu\'{e}s,
{\it Operational Framework for Nonlocality},
\href{http://dx.doi.org/10.1103/PhysRevLett.109.070401}{Phys. Rev. Lett. \textbf{109}, 070401 (2012).}

\bibitem{V14}
J. I. de Vicente, 
{\it On nonlocality as a resource theory and nonlocality measures}, 
\href{https://doi.org/10.1088/1751-8113/47/42/424017}{J. Phys. A: Math. Theor. \textbf{47}, 424017 (2014).}

\bibitem{GA17}
R. Gallego and L. Aolita, 
{\it Nonlocality free wirings and the distinguishability between Bell boxes}, 
\href{http://dx.doi.org/10.1103/PhysRevA.95.032118}{Phys. Rev. A \textbf{95}, 032118 (2017).}

\bibitem{BG15}
F. G. S. L. Brand\~ao and G. Gour,
{\it Reversible Framework for Quantum Resource Theories},
\href{http://dx.doi.org/10.1103/PhysRevLett.115.070503}{Phys. Rev. Lett. \textbf{115}, 070503 (2015);}
\href{https://doi.org/10.1103/PhysRevLett.115.199901}{Phys. Rev. Lett. \textbf{115}, 199901 (2015).}

\bibitem{CFS16}
B. Coecke, T. Fritz, and R. W. Spekkens, 
{\it A mathematical theory of resources}, 
\href{http://dx.doi.org/10.1016/j.ic.2016.02.008}{Inf. Comput. \textbf{250}, 59 (2016).} 

\bibitem{Barrett05b}
J. Barrett, N. Linden, S. Massar, S. Pironio, S. Popescu, and D. Roberts, 
{\it Nonlocal correlations as an information-theoretic resource}, 
\href{https://doi.org/10.1103/PhysRevA.71.022101}{Phys. Rev. A \textbf{71}, 022101 (2005).}

\bibitem{Allcock09}
J. Allcock, N. Brunner, N. Linden, S. Popescu, P. Skrzypczyk, and T. V\'ertesi, 
{\it Closed sets of nonlocal correlations},
\href{https://doi.org/10.1103/PhysRevA.80.062107}{Phys. Rev. A \textbf{80}, 062107 (2009).}

\bibitem{Joshi13}
P. Joshi, M. Horodecki, R. Horodecki, A. Grudka, K. Horodecki, and P. Horodecki, 
{\it No-broadcasting of non-signaling boxes via operations which transform local boxes into local ones}, 
Quantum Info. Comput. \textbf{13}, 567 (2013).

\bibitem{LVN14} 
B. Lang, T. V\'ertesi, and M. Navascu\'es,
{\it Closed sets of correlations: answers from the zoo}, 
\href{https://doi.org/10.1088/1751-8113/47/42/424029}{J. Phys. A: Math. Theor. \textbf{47}, 424029 (2014).}

\bibitem{GA15}
R. Gallego and L. Aolita, 
{\it Resource theory of steering}, 
\href{https://doi.org/10.1103/PhysRevX.5.041008}{Phys. Rev. X \textbf{5}, 041008 (2015).}

\bibitem{vDGG05}
W. van Dam, R. D. Gill, and P. D. Gr\"unwald,
{\it The statistical strength of nonlocality proofs},
\href{https://doi.org/10.1109/TIT.2005.851738}{IEEE Trans. Inf. Theory \textbf{51}, 2812 (2005).}

\bibitem{comment}
In fact, in Ref.\ \cite{GHHHJKW14}, monotonicity of another related quantity is shown, the uniform relative entropy of contextuality, under the same restricted subclass. However, one can show that the uniform variant is not monotonous under more general wirings in $\mathsf{NCW}$, even restricted to trivial (identity) post-processing boxes \cite{GA17}.

\bibitem{BC90}
S. L. Braunstein and C. M. Caves, 
{\it Wringing out better Bell inequalities},
\href{https://doi.org/10.1016/0003-4916(90)90339-P}{Ann. Phys. \textbf{202}, 22 (1990).}

\bibitem{CHSH69}
J. F. Clauser, M. A. Horne, A. Shimony, and R. A. Holt,
{\it Proposed Experiment to Test Local Hidden-Variable Theories},
\href{http://dx.doi.org/10.1103/PhysRevLett.23.880}{Phys. Rev. Lett. \textbf{23}, 880 (1969).}
 
\bibitem{KCBS08}
A. A. Klyachko, M. A. Can, S. Binicio\u{g}lu, and A. S. Shumovsky,
{\it Simple Test for Hidden Variables in Spin-1 Systems},
\href{http://dx.doi.org/10.1103/PhysRevLett.101.020403}{Phys.~Rev.~Lett. \textbf{101}, 020403 (2008).}
 
\bibitem{AQBTC13}
M. Ara\'ujo, M. T. Quintino, C. Budroni, M. Terra Cunha, and A. Cabello,
{\it All noncontextuality inequalities for the $n$-cycle scenario},
\href{http://dx.doi.org/10.1103/PhysRevA.88.022118}{Phys. Rev. \textbf{88}, 022118 (2013).}
 
\bibitem{PR94}
S. Popescu and D. Rohrlich,
{\it Quantum nonlocality as an axiom},
\href{http://dx.doi.org/10.1007/BF02058098}{Found. Phys. \textbf{24}, 379 (1994).}

%%%%%%%%%%%%%%%%%%%%%%%%%%%%%%%%%%%%%%%%%%%%%%%%%%%%%%%%%%%%%%%%%%%

%\bibitem{BKP06}
%J. Barret, A. Kent, and S. Pironio, Phys. Rev. Lett. \textbf{97}, 170409 (2006).
%

% \bibitem{CY15}
% G. Chiribella and X. Yuan,
% Bridging the gap between general probabilistic theories and the device-independent framework for nonlocality and contextuality,
% \href{http://arxiv.org/abs/1504.02395}{arXiv:1504.02395.}

%\bibitem{Tsirelson80}
% B. S. Cirel'son [Tsirelson],
% Quantum generalizations of Bell's inequality,
% \href{http://dx.doi.org/10.1007/BF00417500}{Lett. Math. Phys. \textbf{4}, 93 (1980).}
% 
% \bibitem{Tsirelson93}
% B. S. Cirel'son [Tsirelson],
% Some results and problems on quantum Bell-type inequalities,
% \href{http://www.tau.ac.il/~tsirel/download/hadron.html}{Hadronic J. Supp. \textbf{8}, 329 (1993).}

%%%%%%%%%%%%%%%%%%%%%%%%%%%%%%%%%%%%%%%%%%%%%%%%%%%%%%%%%%%%%%%%%%%

\end{thebibliography}
\end{document}